\documentclass[journal]{IEEEtran}
\usepackage{xcolor}
\usepackage{tcolorbox}
\tcbuselibrary{listings}
\usepackage{enumitem}
\usepackage{amsfonts}
\usepackage{enumitem,kantlipsum}
\usepackage{textcomp}
\usepackage{stfloats}
\usepackage{url}
\usepackage{caption}
\usepackage{graphics}
\usepackage{verbatim}
\usepackage{pifont}
\usepackage{float}
\usepackage{amsmath}
\usepackage{cancel}
\usepackage{amsmath}
\usepackage{graphicx}
\usepackage{wrapfig}
\usepackage{float}
\usepackage{lscape}
\usepackage{ltxtable}
\usepackage{hhline}
\usepackage{caption}
\usepackage{wasysym}
\usepackage{xurl}
\usepackage{rotating}
\usepackage{graphicx}
\usepackage{caption}
\usepackage{array}
\usepackage{listings}
\usepackage{subcaption}
\usepackage{algorithm} 
\usepackage[draft]{hyperref}
\usepackage{color} 
\usepackage[export]{adjustbox}
\usepackage{colortbl}
\usepackage{algpseudocode}
\usepackage{verbatim}
\usepackage{lipsum}
\usepackage{pbox}

\usepackage{tikz}
\usetikzlibrary{shapes,arrows}
\usetikzlibrary{intersections}
\usetikzlibrary{automata, arrows.meta, positioning}
\usetikzlibrary{decorations.pathreplacing}
\usepackage{amsthm}
\usepackage{booktabs}
\usepackage{amsmath}
\usepackage{algorithm}
\usepackage{inputenc}
\usepackage{adjustbox}
\usepackage{url} 
\usepackage{glossaries}

\usepackage{multirow}
\usepackage{color,soul}
\usepackage{lipsum}
\usepackage{mathtools}
\usepackage{subcaption}
\usepackage{cuted}
\usepackage{wasysym}
\usepackage{pgfplotstable}
\usepackage{mdframed}
 \usepackage{mhchem}

\newtheorem{thm}{Theorem}

\newtheorem{lem}{Lemma}

\newtheorem{cor}{Corollary}
\theoremstyle{definition}

\newtheorem{game}{Game}
\theoremstyle{remark}

\usepackage[normalem]{ulem}
\usetikzlibrary{arrows.meta}
\usetikzlibrary{positioning, calc}
\usepackage{smartdiagram}
\definecolor{lemon}{rgb}{1.0, 1.0, 0.13}
\usetikzlibrary{arrows.meta}
\usepackage{smartdiagram}
\usetikzlibrary{positioning, fit, calc, shapes, arrows,trees}

\definecolor{columbiablue}{rgb}{0.61, 0.87, 1.0}

\usepackage{algpseudocode}

\makeatother
\tikzset{%
 thick arrow/.style={
 -{Triangle[angle=120:1pt 1]},
 line width=0.8cm, 
 draw=blue!20
 },
 arrow label/.style={
 text=black,
 align=center
 },
 set mark/.style={
 insert path={
 node [midway, arrow label, node contents=#1]
 }
 }
}

\usepackage{tikz}
\usetikzlibrary{positioning}

\usepackage[cmintegrals]{newtxmath}

\usepackage{hhline}
\definecolor{columbiablue}{rgb}{0.61, 0.87, 1.0}

\tikzset{%
 thick arrow/.style={
 -{Triangle[angle=120:1pt 1]},
 line width=0.8cm, 
 draw=blue!20
 },
 arrow label/.style={
 text=black,
 align=center
 },
 set mark/.style={
 insert path={
 node [midway, arrow label, node contents=#1]
 }
 }
}
\newcommand\deleted{\bgroup\markoverwith{\textcolor{red}{\rule[0.5ex]{2pt}{0.4pt}}}\ULon}

\usetikzlibrary{pgfplots.groupplots}

\newacronym{AES}{AES}{Advanced Encryption Standard}
\newacronym{CCPA}{CCPA}{California Consumer Privacy Act}
\newacronym{CID}{CID}{Content Identifier}
\newacronym{DP}{DP}{Differential Privacy}
\newacronym{DDH}{DDH}{Decisional Diffie-Hellman}

\newacronym{DLP}{DLP}{Discrete Logarithm Problem}
\newacronym{ECDH}{ECDH}{Elliptic Curve Diffie-Hellman}
\newacronym{EHRs}{EHRs}{Electronic Health Records}
\newacronym{GDP}{GDP}{Global Differential Privacy}
\newacronym{GDPR}{GDPR}{General Data Protection Regulation}
\newacronym{IND-CPA}{IND-CPA}{Indistinguishability under Chosen Plaintext Attack}
\newacronym{INT-CT}{INT-CT}{Integrity of Ciphertexts}
\newacronym{IPFS}{IPFS}{InterPlanetary File System}
\newacronym{IoMT}{IoMT}{Internet of Multimedia Things}
\newacronym{IoT}{IoT}{Internet of Things}
\newacronym{LDP}{LDP}{Local Differential Privacy}
\newacronym{MFA}{MFA}{Multi-Factor Authentication}
\newacronym{ML}{ML}{Machine Learning}
\newacronym{OPRF}{OPRF}{Oblivious Pseudorandom Function}
\newacronym{PKI}{PKI}{Public Key Infrastructure}

\newacronym{RTT}{RTT}{Round-Trip Time}
\newacronym{PPT}{PPT}{probabilistic polynomial-time}
\newacronym{PRF}{PRF}{Pseudorandom Function}
\newacronym{QID}{QID}{Quasi-Identifier}
\newacronym{RBA}{RBA}{Risk-Based Adaptive Authentication}
\newacronym{RLAuth}{RLAuth}{Reinforcement Learning-Based Authentication}
\newacronym{ZKPs}{ZKPs}{Zero-Knowledge Proofs}

\begin{document}

\title{Privacy-Enhanced Adaptive Authentication: User Profiling with Privacy Guarantees}

\author{Yaser Baseri,
 Abdelhakim Senhaji Hafid,
 and 
Dimitrios Makrakis
 
\IEEEcompsocitemizethanks{\IEEEcompsocthanksitem Yaser Baseri,  and Abdelhakim Senhaji Hafid are Department of Computer Science and Operations Research, University of Montreal, Canada.
Emails: yaser.baseri@umontreal.ca; ahafid@iro.umontreal.ca \protect
\IEEEcompsocthanksitem Dimitrios Makrakis is with School of Electrical Engineering and Computer Science, University of Ottawa, Canada. E-mail: dmakraki@uottawa.ca\protect} 
\thanks{This work was supported by Natural Sciences and Engineering Research
Council of Canada (NSERC) and Flex Group  (\url{www.flexgroups.com}).}}

\markboth{}%
{Baseri \MakeLowercase{\textit{et al.}}: Privacy-Enhanced Adaptive Authentication: User Profiling with Privacy Guarantees}

\maketitle

\begin{abstract}

User profiling is fundamental to adaptive risk-based authentication but raises significant privacy concerns, particularly regarding the creation of quasi-identifiers that can enable re-identification attacks even on anonymized data. This paper presents a privacy-enhanced adaptive authentication protocol that combines \gls{OPRF}, anonymous tokens, and \gls{DP} to provide robust privacy guarantees while maintaining dynamic risk assessment capabilities.
Our protocol introduces three key innovations: (1) a privacy-preserving risk assessment mechanism that dynamically adjusts authentication requirements without compromising user anonymity, (2) a decentralized profile management system that prevents unauthorized access and re-identification attacks through advanced cryptographic techniques, and (3) the integration of differential privacy mechanisms to minimize the impact of individual data points on overall analysis. Comprehensive privacy guarantees are supported by  security proofs, ensuring confidentiality, integrity, and unlinkability of user data. These features naturally align with data protection regulations such as \gls{GDPR} and \gls{CCPA}, enhancing user trust and ensuring compliance with legal standards.
A thorough performance evaluation demonstrates the protocol's practicality, with manageable computational and communication overheads that enable real-world deployment. This work bridges a critical gap in traditional adaptive authentication systems by integrating privacy protection at the architectural level while preserving the benefits of risk-based authentication. The proposed solution is particularly valuable for sectors handling sensitive data, such as healthcare and finance, where both security and privacy are paramount. Additionally, it contributes to the broader development of secure and privacy-respecting digital ecosystems.
development of secure and privacy-respecting digital ecosystems.
\end{abstract}

\IEEEpeerreviewmaketitle
\section{Introduction}

\IEEEPARstart {T}{he} landscape of user authentication has undergone a significant evolution, transitioning from static methodologies to adaptive and risk-based approaches. While these advancements have enhanced security through \gls{RBA}, they have also underscored the critical role of user profiling~\cite{eke2019survey}, raising significant privacy concerns. Traditional user profiling techniques remain susceptible to re-identification attacks through the exploitation of quasi-identifiers, despite anonymization efforts. This vulnerability presents a fundamental challenge: balancing robust security with user privacy, particularly in sectors handling sensitive data. The implementation of data protection regulations like \gls{GDPR}\cite{gdpr2018general} and \gls{CCPA}\cite{goldman2020introduction} further necessitates robust privacy-preserving solutions.

The risks associated with user profiling are especially evident in sensitive applications, such as e-health, where confidential data like disease types must be securely protected. A seminal U.S. Census study demonstrated that 87\% of the population could be uniquely identified using only ZIP code, sex, and birthdate~\cite{sweeney2002k}. These quasi-identifiers enable re-identification of individuals even from anonymized datasets, highlighting the importance of developing authentication systems that not only enhance security but also safeguard user privacy against emerging threats~\cite{HIRSCHPRUNG2021101564}.

Privacy-preserving \gls{RBA} and user profiling have wide-ranging applications across various domains, each requiring a delicate balance between robust security measures and user privacy. Financial institutions handle highly sensitive information like account details and transaction history, necessitating secure access while protecting user data from profiling that could enable targeted scams or marketing~\cite{FFIEC2021}. In healthcare, \gls{EHRs} contain extremely sensitive personal data, requiring authentication methods that ensure secure access for authorized personnel while anonymizing user profiles~\cite{basil2022health}. Government and military systems demand robust security measures~\cite{whitehouse2023cybersecurity} while protecting sensitive governmental and military data~\cite{force2017security}, particularly for applications related to social security, taxes, and law enforcement. The significance of privacy-preserving techniques extends to e-commerce, where the increasing prevalence of online transactions necessitates protection against data breaches and unauthorized profiling~\cite{pooranian2021online}. These principles apply broadly to social media platforms, messaging apps, cloud storage services, and any system handling personal or sensitive data.

While \gls{RBA} offers substantial security benefits by dynamically adjusting authentication measures based on user behavior, context, and risk scores, it often compromises user privacy. This fundamental tension between security and privacy underscores the importance of addressing privacy concerns in authentication systems. Our work proposes a novel privacy-preserving risk-based adaptive authentication protocol to counter vulnerabilities in existing systems, thereby mitigating privacy concerns and fostering increased user trust and legal compliance. Leveraging cryptographic methods and privacy-enhancing technologies, our methodology preserves the efficacy of authentication mechanisms without compromising user anonymity. Our approach aligns with evolving legal frameworks protecting user data, ensuring compliance with data protection  regulations, and contributes to the broader development of secure and privacy-respecting digital ecosystems.

The main contributions of this paper can be summarized as follows:

\begin{itemize}
    \item \textbf{Privacy-Preserving Risk-Based Authentication:} We propose a novel framework using \gls{OPRF}, anonymous tokens, and \gls{DP} to enable risk-based  authentication while preserving the anonymity of users. This approach ensures unlinkability and prevents user   tracking, and inference attacks, aligning with data protection regulations.

    \item \textbf{Formal Security and Privacy Proofs:} We provide rigorous security proofs, demonstrating that our protocol satisfies confidentiality, integrity, and unlinkability, and resists re-identification, tracking, and inference attacks under standard cryptographic assumptions.

    \item \textbf{Scalability, Efficiency and Seamless Integration:} Our protocol's low computational and communication overhead ensures feasibility for real-world deployment. Designed as a privacy-enhancing layer, it integrates with existing RBA frameworks, allowing organizations to adopt privacy-preserving mechanisms without disrupting current infrastructure.
\end{itemize}

The remainder of this paper is structured as follows: Section \ref{sec:related} reviews literature on adaptive authentication and associated privacy challenges. Section \ref{sec:UP} presents a taxonomy of user profiling methods. Section \ref{sec:preliminaries} provides necessary preliminaries. Section \ref{sec:models} describes system and security models. Section \ref{sec:proposed} details our proposed approach. Section \ref{sec:security} analyzes security and privacy guarantees. Section \ref{sec:performance} evaluates performance. Section \ref{sec:conclusion} concludes with key findings and contributions.

\section{Related Works}\label{sec:related}

Privacy-preserving  \gls{RBA} and user profiling are critical areas of research, especially with the growing need to protect user data in various applications. Several studies have proposed different approaches to achieve these objectives. This section reviews the existing literature on privacy-preserving \gls{RBA} and user profiling.

\subsection{Risk-Based Authentication and \gls{ML}}

 \gls{RBA} dynamically adjusts security measures based on the perceived risk of a login attempt, aiming to balance security with user experience. By leveraging factors such as knowledge, biometrics, behavior, context, or tokens,  \gls{RBA} systems assess the likelihood of a fraudulent attempt and apply corresponding authentication challenges.

Several studies have explored \gls{RBA} mechanisms. Sepczuk and Kotulski~\cite{sepczuk2018new} propose a model incorporating contextual data (e.g., user security experience and service type) to refine risk assessment and authentication decisions. Papaioannou et al.~\cite{papaioannou2022toward} focus on mobile device authentication in border control, using novelty detection to identify anomalies. 
Picard and Pierre~\cite{picard2023rlauth} propose \gls{RLAuth}, a deep reinforcement learning-based system that dynamically adjusts authentication challenges based on assessed risk. Singh et al.~\cite{singh2022resilient} introduce a risk-based framework leveraging \gls{ML} to classify authentication contexts and adapt security measures accordingly. Unsel et al.~\cite{unsel2023risk} provide a practical \gls{RBA} implementation for OpenStack, demonstrating its feasibility for real-world deployment. Progonov et al.~\cite{progonov2022behavior} utilize user behavior patterns as a non-intrusive authentication method and propose BehaviorID, a system that continuously monitors user behavior for real-time authentication, achieving high accuracy and robustness against spoofing attacks. Gupta et al.~\cite{gupta2019driverauth} propose a biometric driver authentication system for ride-sharing, integrating swipe gestures, voice-prints, and face images to enhance security and usability. Yang et al.~\cite{yang2024cross} develop a server-orchestrated federated learning framework for cross-device keystroke authentication, extracting device-agnostic keystroke dynamics features and optimizing an authentication model through auxiliary learning with Gaussian-based user representations.



While these approaches highlight the potential of \gls{RBA}, they face inherent challenges in balancing security, privacy, and usability. Server-centric approaches, such as those in \cite{sepczuk2018new, papaioannou2022toward, singh2022resilient, unsel2023risk, gupta2019driverauth}, rely on centralized data aggregation, raising concerns over privacy breaches, profiling, and misuse. Even with anonymization, continuous behavioral monitoring can enable user re-identification through sensitive attributes such as location data and device fingerprints. In contrast, device-centric approaches, such as those in \cite{picard2023rlauth, progonov2022behavior}, mitigate centralized risks by keeping data on the user's device. However, they face computational constraints, cross-device compatibility issues, and require extensive data collection, particularly during the cold-start phase. Moreover, they do not inherently provide strong privacy protections, as local adversaries can still perform profiling and inference attacks over time. 
Hybrid approaches, such as \cite{yang2024cross,fereidouni2024f}, leverage federated learning to balance the trade-offs between server- and device-centric approaches by distributing computation. However, they still inherit privacy risks, as aggregated model updates can be exploited for sensitive user behavior inference through techniques like membership inference and model inversion.


\subsection{Privacy-Preserving Authentication}
Privacy-preserving authentication protocols employ diverse techniques to ensure secure and confidential user interactions, including (1) \emph{Anonymous Tokens and \gls{ZKPs}}~\cite{silde2022anonymous,mir2020damfa},  (2) \emph{Encryption Schemes}~\cite{domingo2015flexible,baig2023privacy},  (3) \emph{Differential Privacy}~\cite{usman2019paal,chamikara2020privacy}, and  (4) \emph{Blockchain-Based Techniques}~\cite{kim2022multi}.  

Anonymous tokens and \gls{ZKPs} prioritize static identity protection, yet often lack dynamic adaptability. Knijnenburg \cite{knijnenburg2017privacy} highlights the need for context-aware privacy mechanisms, while Silde and Strand \cite{silde2022anonymous} introduce tokens with public metadata for private contact tracing. Mir et al. \cite{mir2020damfa} propose DAMFA, a decentralized multi-factor scheme using blockchain and \gls{ZKPs}, eliminating trusted third parties but lacking risk-based adjustments. Encryption-based approaches, such as Domingo-Ferrer et al.’s set intersection protocol \cite{domingo2015flexible} and Baig et al.’s homomorphic encryption for continuous authentication \cite{baig2023privacy}, ensure data confidentiality but focus on static authentication without dynamic risk assessment or policy enforcement.
Differential privacy techniques, like Usman et al.’s edge-computing framework for \gls{IoMT} \cite{usman2019paal} and Chamikara et al.’s Eigenface obfuscation \cite{chamikara2020privacy}, balance privacy and utility but face scalability challenges and lack explicit risk-adaptive mechanisms. Blockchain-based methods, exemplified by Kim et al.’s decentralized biometric \gls{MFA} \cite{kim2022multi}, enhance data integrity but struggle with scalability and dynamic risk adaptation.

While these methods are effective for anonymity, data secrecy, or re-identification resistance, they lack inherent support for adaptive risk-based authentication. Encryption and blockchain approaches often overlook dynamic risk assessment, while differential privacy and \gls{ZKPs} do not inherently adjust authentication strength based on risk, a critical requirement for \gls{RBA}.

\subsection{Summary of Research Gaps and  Challenges}
While existing research on \gls{RBA} and privacy-preserving authentication demonstrates their effectiveness, a significant gap remains in providing privacy-preserving risk-based adaptive authentication that effectively mitigates the risk of user re-identification (de-anonymization). This gap is further enhanced when considering quasi-identifiers. To the best of our knowledge, none of the existing risk-based adaptive authentication systems adequately {addresses} privacy concerns, while privacy-preserving authentication methods often lack the adaptability required for risk-based decision-making. Our proposed approach
addresses this challenge by offering robust privacy guarantees while maintaining adaptive authentication capabilities. By leveraging \gls{OPRF}, anonymous tokens, \gls{DP}, and \gls{IPFS}, our solution enhances privacy, prevents user tracking, and ensures secure data storage. Our approach is designed for practical implementation and outperforms existing methods in terms of privacy and security. 


\section{User Profiling and \gls{RBA}}\label{sec:UP}

\gls{RBA}, dynamically adjusts authentication requirements based on perceived login attempt risk.  It enhances traditional methods by incorporating risk evaluation without requiring extra user interaction, applying lightweight measures for low-risk access and stronger verification (e.g., OTPs, biometrics) for suspicious behavior \cite{unsel2023risk,wiefling2022rba}. This proactive, "security by design" approach reinforces account protection, even with valid credentials.

\subsection{User Profiling and \gls{RBA}: Mechanism}\label{risk-evaluation}

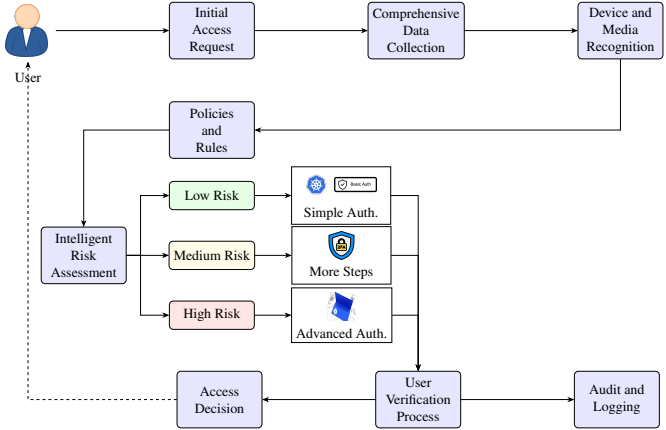
\begin{figure}
    \centering
    \Large
    \resizebox{\linewidth}{!}{
    \begin{tikzpicture}[scale=1,
        node distance=3cm and 4cm,
        every node/.style={align=center},
        arrow/.style={thick,->,>=Stealth}
        ]

        \node (user) [draw=none, fill=none] {
            \includegraphics[width=1.5cm]{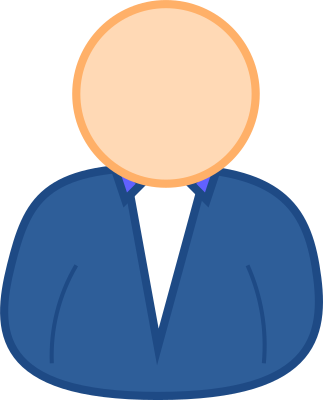}
        };
        \node [below=0.2cm of user] {User};

        \node (request) [right=of user, draw, minimum width=3cm, minimum height=2cm, align=center,fill=blue!10,rectangle, draw, rounded corners, text centered] {Initial \\Access\\ Request};

        \node (data) [right=of request, draw, minimum width=3cm, minimum height=2cm, align=center,fill=blue!10,rectangle, draw, rounded corners, text centered] {Comprehensive\\Data\\Collection};

        \node (device) [right=of data, draw, minimum width=3cm, minimum height=2cm, align=center,fill=blue!10,rectangle, draw, rounded corners, text centered] {Device and\\Media\\Recognition};

        \node (policies) [below=of request, yshift=1.5cm, draw, minimum width=3cm, minimum height=2cm, align=center,fill=blue!10,rectangle, draw, rounded corners, text centered] {Policies\\and\\Rules};

        \node (LowRisk) [below=0.7cm of policies, draw, minimum width=3cm, minimum height=1cm, align=center,fill=green!10,rectangle, draw, rounded corners, text centered] {Low Risk};
        \node (MediumRisk) [below=0.8cm of LowRisk, draw, minimum width=3cm, minimum height=1cm, align=center,fill=yellow!10,rectangle, draw, rounded corners, text centered, yshift=-0.4cm] {Medium Risk};
        \node (HighRisk) [below=0.8cm of MediumRisk, draw, minimum width=3cm, minimum height=1cm, align=center,fill=red!10,rectangle, draw, rounded corners, text centered, yshift=-0.4cm] {High Risk};

        \node (assessment) [left=of MediumRisk, xshift=2.5cm, draw, minimum width=3cm, minimum height=2cm, align=center,fill=blue!10,rectangle, draw, rounded corners, text centered] {Intelligent\\Risk\\Assessment};

        \node (SimpleAuth) [right=of LowRisk, draw, align=center, minimum width=3.5cm, xshift=-2.7cm] {
            \includegraphics[height=1cm, width=2.5cm,trim=100 120 90 150,clip]{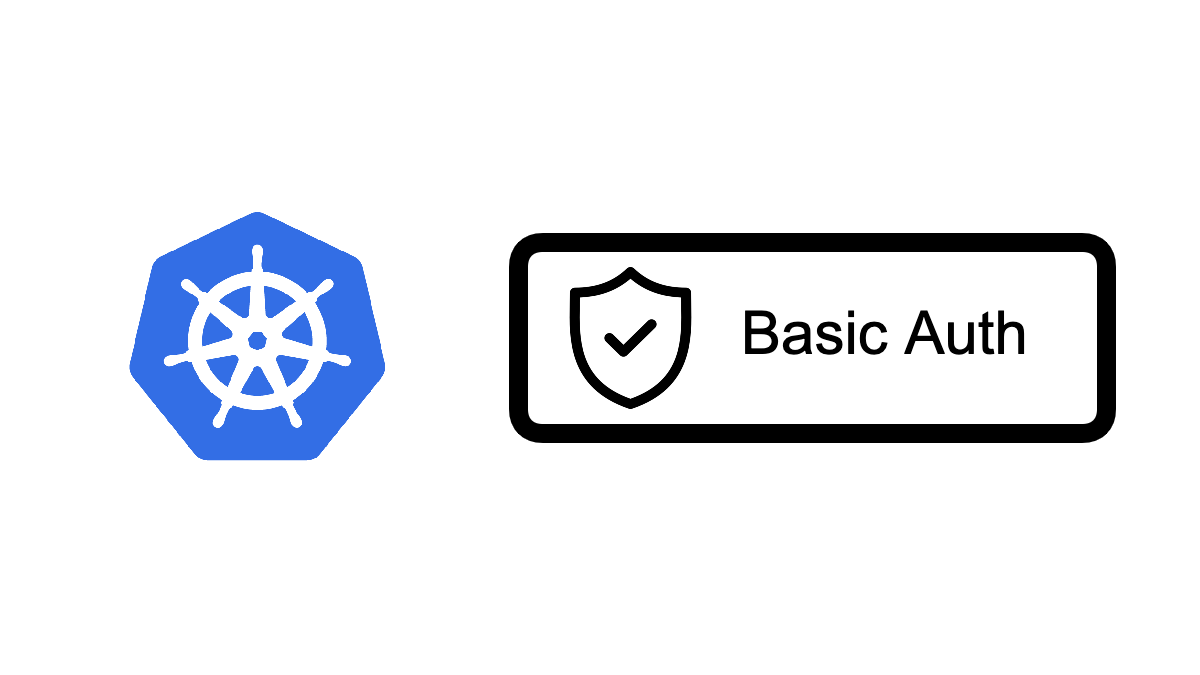} \\ Simple Auth.
        };
        \node (MoreSteps) [right=of MediumRisk, align=center, minimum width=3.5cm, xshift=-2.7cm, draw] {
            \includegraphics[height=1cm, width=1cm,trim=0 0 0 0,clip]{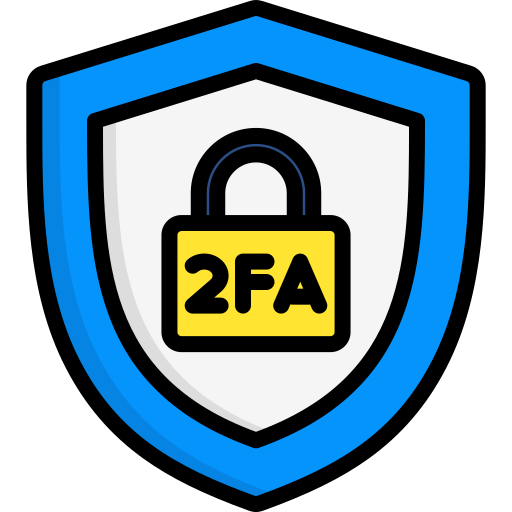} \\ More Steps
        };
        \node (AdvancedAuth) [right=of HighRisk, align=center, minimum width=3.5cm, xshift=-2.7cm, draw] {
            \includegraphics[height=1cm, width=1cm,trim=50 50 50 50,clip]{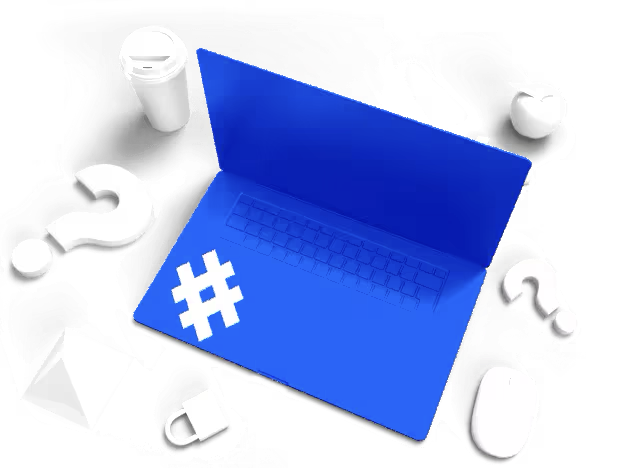} \\ Advanced Auth.
        };

        \node (UserVerification) [below=1cm of AdvancedAuth, xshift=2.7cm, draw, minimum width=3cm, minimum height=2cm, align=center, fill=blue!10,rectangle, draw, rounded corners, text centered] {User \\ Verification \\ Process};
        \node (AccessDecision) [left=of UserVerification, draw, minimum width=3cm, minimum height=2cm, align=center, fill=blue!10,rectangle, draw, rounded corners, text centered] {Access \\Decision };

        \node (logging) [right=of UserVerification, draw, minimum width=3cm, minimum height=2cm, align=center, fill=blue!10,rectangle, draw, rounded corners, text centered] {Audit and\\Logging};

        \draw[arrow] (user) -- (request.west);
        \draw[arrow] (request.east) -- (data);
        \draw[arrow] (data.east) -- (device);
        \draw[arrow] (device) |- (policies);
        \draw[arrow] (policies) -| (assessment);
        \draw[arrow] (assessment.east) -- ++(.5cm, 0)|- (LowRisk.west);
        \draw[arrow] (assessment.east) -- (MediumRisk);
        \draw[arrow] (assessment.east) -- ++(.5cm, 0)|- (HighRisk.west);
        \draw[arrow] (LowRisk.east) -- (SimpleAuth);
        \draw[arrow] (MediumRisk.east) -- (MoreSteps);
        \draw[arrow] (HighRisk.east) -- (AdvancedAuth);
        \draw[arrow] (SimpleAuth) -| (UserVerification); 
        \draw[arrow] (MoreSteps) -| (UserVerification); 
        \draw[arrow] (AdvancedAuth) -| (UserVerification); 
        \draw[arrow] (UserVerification) -- (AccessDecision);

        \draw[arrow, dashed] (AccessDecision.west) -| (user.south);
        \draw[arrow] (UserVerification) -- (logging);

    \end{tikzpicture}}
    \caption{Risk Based  Authentication Mechanism}
       \vspace{-0.4cm}
    \label{fig:user_transaction_flowchart}
\end{figure}

\gls{RBA} evaluates login risk using features from five categories: knowledge-based (e.g., passwords, PINs), biometric (e.g., fingerprints, facial recognition), behavioral (e.g., typing patterns, mouse movements), contextual (e.g., geolocation, device type), and interaction-based (e.g., session frequency, navigation flow).
 It categorizes risk (e.g., low, medium, high) and applies authentication accordingly, ranging from passwords to multifactor or biometric checks.
As illustrated in Figure \ref{fig:user_transaction_flowchart}, this process includes collecting and preprocessing user data (e.g., IP address, location, OS, keystrokes), applying organizational rules, and determining appropriate authentication steps.  The user is then verified and the event is recorded for audit and compliance.
\gls{RBA} uses \gls{ML} to dynamically assess risk. Supervised models (e.g., logistic regression, decision trees) classify transactions by risk \cite{paltrinieri2019learning}; unsupervised methods (e.g., isolation forests) detect anomalies \cite{ashtari2022comparative}; and reinforcement learning optimizes policy over time \cite{guan2021reinforcement,hazratifard2022using}. Feature engineering (e.g., typing speed, transaction patterns) refines detection and enhances behavioral profiling \cite{lopez2021feature,djosic2020machine}.  \gls{RBA} systems integrate with identity and access management platforms to support risk-aware access across sectors like banking, e-commerce, and healthcare \cite{knoblauch2023towards,li2022rebuilding}, minimizing user friction while mitigating evolving cyber threats. Core components include: (i) a risk evaluation engine; (ii) device/media recognition; (iii) dynamic \gls{MFA} selection; (iv) re-authentication; (v) contextual/behavioral analysis; and (vi) audit logging.

\subsection{User Profiling and \gls{RBA}: Privacy Concerns and De-anonymization}

User profiling aggregates identifiers, quasi-identifiers (QIDs), and sensitive features.  Even absent direct identifiers, QIDs (e.g., age, ZIP, gender) enable re-identification when linked to auxiliary datasets \cite{Ron2021Privacy}, motivating privacy models like $k$-anonymity, $l$-diversity, and $t$-closeness \cite{sei2019anonymization,baseri2024statistical}.  Profiling features include explicit (e.g., name, ZIP), behavioral (e.g., browsing activity, keystrokes), inferred (e.g., social ties, geolocation, session patterns), and derived (e.g., psychographic segmentation) data.  Aggregating these increases de-anonymization risk, especially across fragmented, cross-platform profiles.  Robust privacy-preserving mechanisms are thus essential in \gls{RBA} to ensure secure and ethical deployment.

\section{Preliminaries}\label{sec:preliminaries}

This section reviews the core privacy-preserving building blocks—\gls{OPRF}, anonymous tokens, and \gls{DP}—that underpin our protocol.

\subsection{Oblivious Pseudorandom Function (OPRF)}
An \gls{OPRF} enables a server to evaluate a PRF on a user’s input without learning the input, while the user learns only the output~\cite{kolesnikov2017practical}. Let $\text{PRF}_k(x)$ denote a PRF keyed by $k$. The protocol involves:
\begin{itemize}
    \item $\text{Setup}(1^\lambda) \to k$: Server generates a secret key $k$.
    \item $\text{Query}(x, k) \to y$: Server computes $y = \text{PRF}_k(x)$ obliviously.
\end{itemize}

In our protocol, the user hashes their credentials as \( h = H(U \parallel P) \), and the server evaluates the OPRF by computing \( F_k(h) = h^x \bmod q \), where \( x \) is the server’s secret key. This enables credential validation without revealing the user’s input or the server’s key.
\subsection{Anonymous Tokens}
Anonymous tokens provide unlinkability across sessions by digitally signing \gls{OPRF} outputs without revealing user identity.
\begin{itemize}
    \item $\text{TokenGen}(SK, m) \to T$: Server signs message $m = F_k(h)$ to generate token $T = (m, \text{Sig}_{SK}(m))$.
    \item $\text{TokenVerify}(PK, T, m) \to \{0,1\}$: Verifier uses public key $PK$ to validate $T$.
\end{itemize}
Unlinkability ensures that session tokens from the same user cannot be correlated cross sessions. Token integrity is guaranteed through signature verification.

\subsection{Differential Privacy (DP)}

\gls{DP} is a formal framework that ensures the output of a computation does not reveal sensitive information about any individual data point. A randomized algorithm $\mathcal{A}$ is $\epsilon$-differentially private if, for all adjacent datasets $D$ and $D'$:
\[
\Pr[\mathcal{A}(D) \in S] \leq e^\epsilon \Pr[\mathcal{A}(D') \in S],
\]
where $\epsilon \geq 0$ quantifies the privacy loss, with smaller values offering stronger privacy guarantees.

To enforce \gls{DP}, we apply the Laplace mechanism~\cite{kumar2024differential}, which perturbs both the user’s stored profile features \( \mathbf{X}_\text{Profile} \) and real-time authentication features \( \mathbf{X}_\text{Live} \) prior to risk evaluation:
\(
\mathbf{X}' = \mathbf{X} + \text{Lap}\left(\frac{\Delta f}{\epsilon}\right),
\)
where \( \mathbf{X} \in \{\mathbf{X}_\text{Profile}, \mathbf{X}_\text{Live} \} \), \( \Delta f \) is the global sensitivity of the risk scoring function \( f \), and \( \text{Lap}(\cdot) \) denotes the Laplace distribution.
The server then computes the risk score using the perturbed feature vectors:
\[
R = f(\mathbf{X}'_\text{Profile}, \mathbf{X}'_\text{Live}).
\]
This design mitigates profiling and re-identification risks, ensuring that authentication decisions do not inadvertently expose sensitive user information.

\section{System and Security Models}\label{sec:models}

Our architecture consists of client-side and server-side components supporting privacy-preserving user profiling and risk-based adaptive authentication. The client application extracts features, applies \gls{DP}, encrypts profile data using a symmetric key, and communicates securely with the server. Encrypted profiles are stored locally, with optional backups in decentralized storage via \gls{IPFS}~\cite{medina2024ami}. The server processes authentication requests, computes risk scores on the received private data, and adjusts authentication requirements accordingly.

\begin{figure}[ht]
    \centering
    \resizebox{\linewidth}{!}{
    \begin{tikzpicture}[scale=1, every node/.style={scale=0.8}, node distance=3cm]
        \node (user) [draw=none, fill=none] {\begin{minipage}{0.15\linewidth}\centering
            \includegraphics[width=1cm]{user.png}\\ User \end{minipage}};
        \node [scale=0.9,draw,fill=blue!10, rounded corners,minimum height=4.4cm,minimum width=0.4\linewidth, below=0.2cm of user, align=center,xshift=0] (userbox) {
            \begin{minipage}[t]{0.5\linewidth}
            \vspace{0.6cm}
                \begin{itemize}
                    \item Submit Credentials,
                    \item Interact with Client Application.\vspace{2.3cm}
                \end{itemize}  
            \end{minipage}};

        \node[align=center, right=of user] (client) {\begin{minipage}{0.23\linewidth}\centering
            \includegraphics[width=1cm]{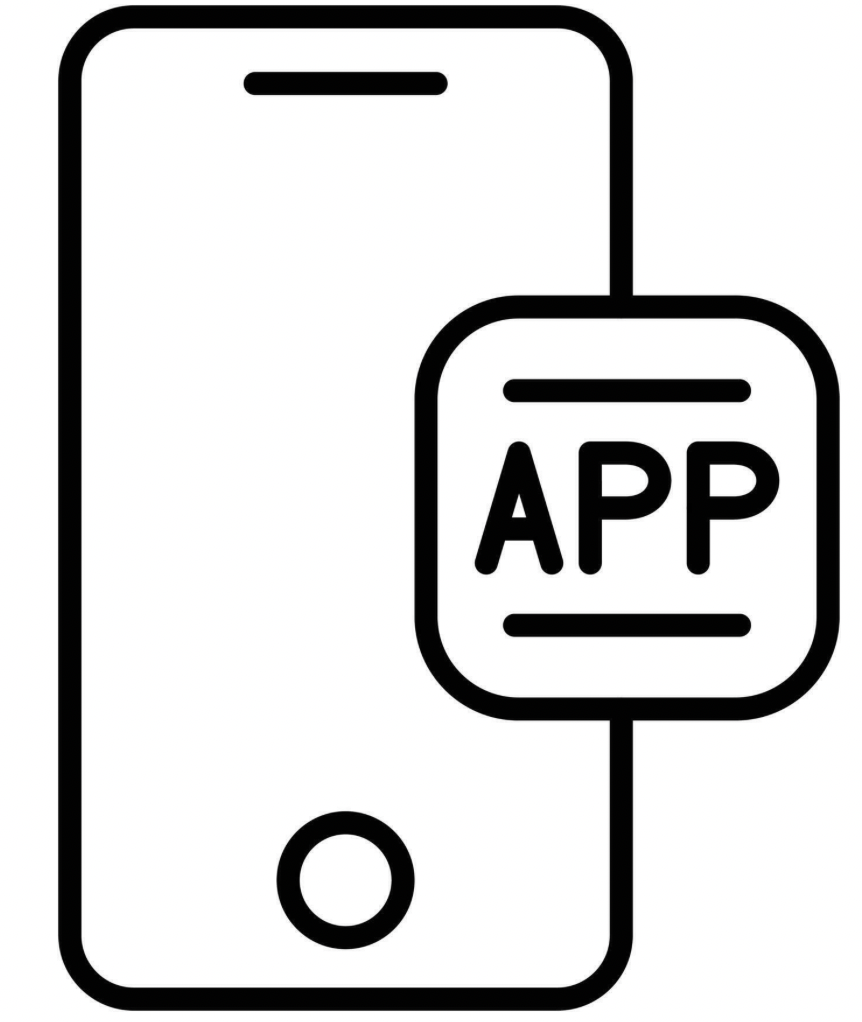}\\ Client App. \end{minipage}};
        \node [scale=0.9,draw,fill=blue!10, rounded corners,minimum height=4.4cm,minimum width=0.4\linewidth, below=0.22cm of client, align=center,xshift=0.5cm] (clientbox) { 
            \begin{minipage}{0.5\linewidth}
                \begin{itemize}
                    \item Encrypt/Decrypt Profile Data,
                    \item Extract Features,
                    \item Apply \gls{DP},
                    \item Secure Communication with Server,
                    \item Backup Encrypted Profile on \gls{IPFS}.
                \end{itemize}  
            \end{minipage}};

        \node[align=center, right=of client, xshift=0.7cm] (server) {\begin{minipage}{0.25\linewidth}\centering
            \includegraphics[width=1cm]{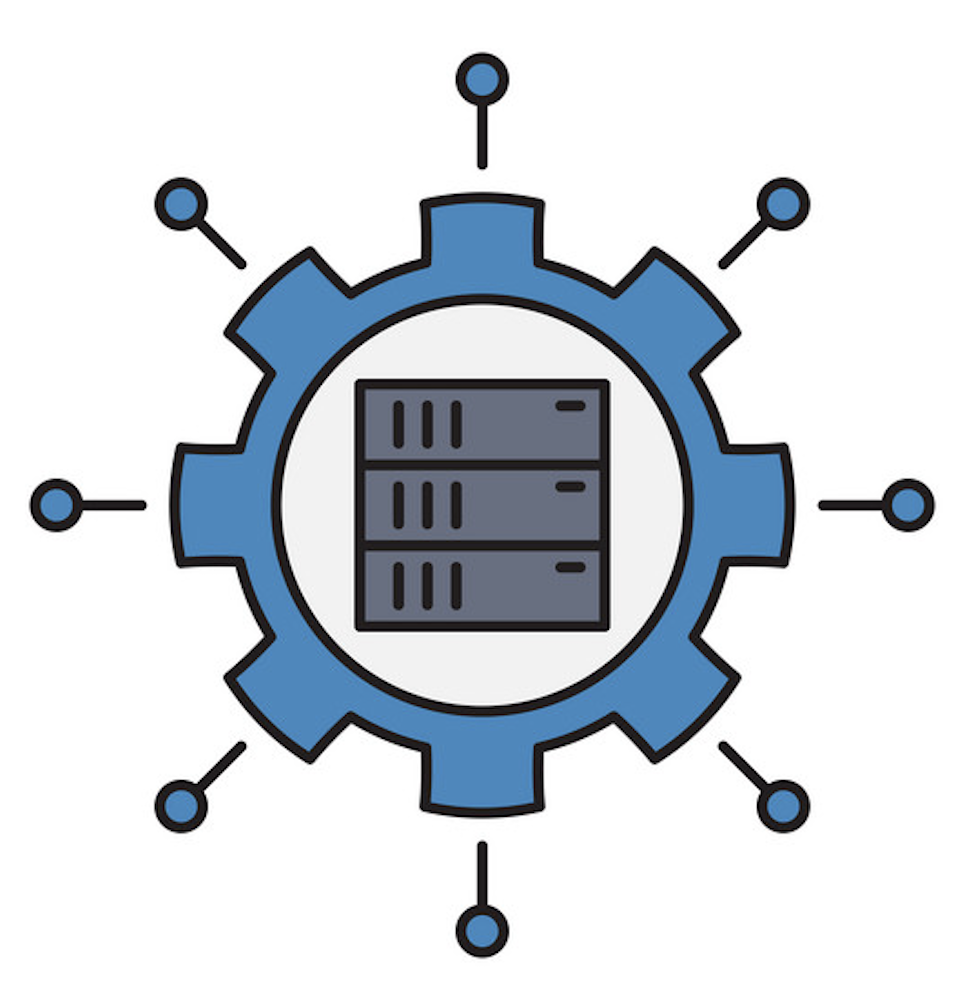}\\ Server \end{minipage}};
        \node [scale=0.9,draw,fill=blue!10, rounded corners,minimum height=4.4cm,minimum width=0.4\linewidth, below=0.3cm of server, align=center,xshift=0cm] (serverbox) {
            \begin{minipage}{0.5\linewidth}
                \begin{itemize}
                    \item Handle Requests,
                    \item Authenticate Users,
                    \item Compute \gls{OPRF},
                    \item Verify Tokens,
                    \item Compute Risk Scores,
                    \item Adjust Auth. Methods.
                \end{itemize}  
            \end{minipage}};

        \node[align=center, above=of client,yshift=-2cm] (ipfs) {\begin{minipage}{0.25\linewidth}\centering
            \includegraphics[width=1cm]{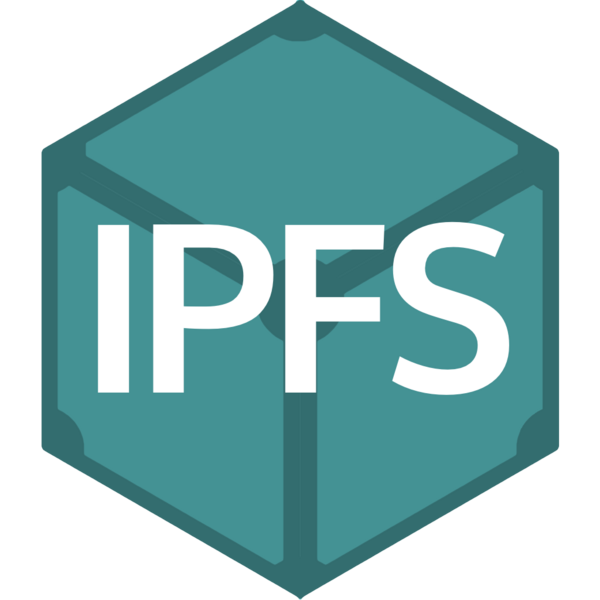}\\ IPFS \end{minipage}};
        \node [scale=0.9,draw,fill=blue!10, rounded corners,minimum height=2cm,minimum width=0.4\linewidth, above=0.2cm of ipfs, align=center,xshift=0] (ipfsbox) {
            \begin{minipage}{0.5\linewidth}
                \begin{itemize}
                    \item Store Encrypted Profile,
                    \item Provide  Content
Identifier (CID),
                    \item Ensure Data Availability.
                \end{itemize}  
            \end{minipage}};

        \draw[->] (user.east) -- ++(0.5,0) |- (client.west) node[midway,above,xshift=1.5cm] {Submit Credentials};

        \draw[->] ($(client.east) + (0,0.6)$)  |- ($(server.west) + (0,0.6)$) node[midway,above,xshift=2.2cm] {Auth. Request, \gls{OPRF} Input};

        \draw[->] ($(server.west) + (0,0.2)$)  |- ($(client.east) + (0,0.2)$) node[midway,above,xshift=-2.3cm] {Auth. Response, Token};

        \draw[->] ($(client.east) - (0,0.2)$)  |- ($(server.west) - (0,0.2)$) node[midway,above,xshift=2.2cm] {Differentially Private Features};

        \draw[->] ($(server.west) - (0,0.6)$)  |- ($(client.east) - (0,0.6)$) node[midway,above,xshift=-2.2cm] {Risk Score, Auth. Adjustment};
        \draw[<->] (client.north) |- (ipfs.south);
        \node[rotate=90, above=of client, yshift=0cm, xshift=-2.8cm] {Backup};
 
    \end{tikzpicture}}
    \caption{System Architecture of the Proposed Protocol}
    \vspace{-0.4cm}
\end{figure}

Data confidentiality and integrity are enforced through HTTPS and multiple cryptographic mechanisms. \gls{OPRF} ensures credentials are validated without revealing inputs. Anonymous tokens provide session unlinkability, and \gls{DP} adds calibrated noise to extracted features before transmission, mitigating profiling risks. Symmetric encryption secures profile data at rest, including backups on \gls{IPFS}.

Our system assumes an honest-but-curious server that follows the protocol but may attempt inference. The server’s private key is securely stored and accessible only to trusted components. The client securely stores its symmetric key and runs in a trusted environment, correctly executing cryptographic operations, feature extraction, and data handling. These assumptions uphold the protocol's security and privacy guarantees.

\section{Proposed Protocol}\label{sec:proposed}

In this section, we present our protocol for privacy-preserving risk-based adaptive authentication. It integrates \gls{OPRF} for secure credential validation, anonymous tokens for unlinkable sessions, and  local differential privacy to protect user features against re-identification. To mitigate aggregation risks and support recoverability, encrypted user profiles are optionally backed up via decentralized \gls{IPFS} storage. This design offers strong privacy guarantees while preserving the adaptive nature of RBA. The protocol consists of the following phases:

\subsection{Setup Phase}

The setup phase establishes the foundational elements of our secure authentication protocol. This phase involves generating the necessary cryptographic keys and issuing anonymous tokens for secure and privacy-preserving authentication. This phase includes the following steps:

\begin{enumerate}[wide, font=\bfseries, labelwidth=!, labelindent=0pt, label=\textbf{Step} \arabic*.]
 \item \textbf{Key Generation:} 
 In this step, both the client application and the server generate the necessary cryptographic keys to secure subsequent communications and data exchanges.
\begin{itemize}
    \item The server selects a private key $ x \in \mathbb{Z}_q $ and computes the public key $ y = g^x \mod q $, where $ g $ is a generator of a cyclic group of prime order $ q $.
\item The client application selects a private key \( a \in \mathbb{Z}_q \) and computes the public key \( A = g^a \mod q \). It also generates a symmetric encryption key \( S_u \) using a secure key generation algorithm. This key is used later for the encryption and decryption of the user's profile data.

\end{itemize}

 \item \textbf{Profile Feature Engineering and Encryption:}
The client application extracts the required features $ \mathbf{X}_\text{Profile} = (x_1, x_2, \ldots, x_n) $ during registration, which will be considered as part of the user's profile. It then encrypts $ \mathbf{X}_\text{Profile} $ using symmetric key cryptography like \gls{AES} and symmetric key $ S_u $:
 \[
 E_{S_u}(\mathbf{X}_\text{Profile}) = \text{\gls{AES}}(S_u, \mathbf{X}_\text{Profile})
 \]
 and stores $ E_{S_u}(\mathbf{X}_\text{Profile}) $ locally.

 \item \textbf{Profile Backup in \gls{IPFS}:}
 The client uploads the encrypted profile $ E_{S_u}(\mathbf{X}_\text{Profile}) $ to \gls{IPFS} and receives a \gls{CID} for the encrypted data. The \gls{CID} is stored locally on the client device.

\item \textbf{Anonymous Token Issuance:} 
This step involves creating and distributing anonymous tokens that enable secure, unlinkable user authentication. In this process:
\begin{itemize}
    \item {The client application prepares the input for the \gls{OPRF} by concatenating the username $U$ and password $P$ and hashing them using a cryptographic hash function $H$. This hashed value $H(\text{U} \parallel \text{P})$ serves as the input for the \gls{OPRF}, ensuring that the token is derived directly from the user's credentials. The security of this process is guaranteed by the properties of the \gls{OPRF}, which ensures that only a valid combination of $U$ and $P$ can produce the correct token, thus verifying the user's identity. The client then selects a blinding factor $b \in \mathbb{Z}_q$ and computes the blinded input for the \gls{OPRF}:}
    \[
    m' = H(\text{U} \parallel \text{P}) \cdot g^b \mod q
    \]
    where $H$ is a cryptographic hash function modeled as a random oracle. The client application then sends $m'$ to the server.

    \item The server computes the \gls{OPRF} evaluation $F_x(m') = (m')^x \mod q$ and sends it to the client.

    \item The client computes $k = (g^x)^b \mod q$ using the server's public key $y = g^x$ and unblinds the \gls{OPRF} evaluation to compute the authentication token as follows:
    \begin{align*}
        F_x(H(\text{U} \parallel \text{P})) &= \frac{F_x(m')}{k} \mod q\\
        &= H(\text{U} \parallel \text{P})^x \mod q
    \end{align*}
\end{itemize}
\end{enumerate}
\subsection{Authentication Phase}
The authentication phase requires the user to authenticate to the server. This is achieved by the server verifying an anonymous token provided by the user, which was issued during the registration phase and is cryptographically linked to the user's credentials. By verifying the token, the server confirms that the user possesses the correct credentials without needing to know them directly. When a session is initiated, an anonymous session token is generated by combining the anonymous token with session-specific data. This integration ensures that the session token is unique to each session, thus safeguarding user privacy and preventing token correlation across sessions. This approach mitigates replay attacks and validates the authentication process, with the secure computation of the session token maintaining confidentiality and robustness against various security threats. The authentication process is carried out through the following steps:



\begin{enumerate}[wide, font=\bfseries, labelwidth=!, labelindent=0pt, label=\textbf{Step} \arabic*.]
\item \textbf{Session Establishment:} 
The client application selects a random number  $ t \in \mathbb{Z}_q $, computes the blinded \gls{OPRF} evaluation $ F_x(H(\text{U} \parallel \text{P}))^{t}  \mod q$ and blinded hash $ H(\text{U} \parallel \text{P})^{t} \mod q $. Then selects a session ID:
\[
\text{SessionID} = \text{Hash}(\text{random bytes})
\]
and blinds it by selecting a blinding factor $ b' \in \mathbb{Z}_q $:
\[
\text{SessionID}' = \text{SessionID} \cdot g^{b'} \mod q
\]
The client application then sends $  F_x(H(\text{U} \parallel \text{P}))^t, H(\text{U} \parallel \text{P})^t, \text{SessionID}') $  to the server.

\item \textbf{Anonymous Session Token Issuance:} 
The server issues a session authentication token corresponding to the user’s authentication request. In this process:
\begin{itemize}
    \item Upon receiving the above message, the server verifies if:
    \[
    F_x(H(\text{U} \parallel \text{P}))^t \stackrel{?}{=} (H(\text{U} \parallel \text{P})^t)^x \mod q
    \] If the equation holds, the server generates  blind token $T'$ by combining its signature on the blinded session ID with the blinded \gls{OPRF} evaluation:
    \[
    T' = F_x(H(\text{U} \parallel \text{P}))^t \cdot (\text{SessionID}')^x \mod q
    \]
    Since $ F_x(H(\text{U} \parallel \text{P})) = H(\text{U} \parallel \text{P})^x $, we have:
    \[
    T' = H(\text{U} \parallel \text{P})^{tx} \cdot \text{SessionID}^x \cdot (g^{x})^{b'} \mod q
    \]. The server then sends $ T' $ to the client.

    \item Client application computes the final token $ T $ by removing the blinding factor on $ T' $:
    \[
    T = T' / (g^x)^{b'} \mod q
    \]
    Since $ y = g^x $ is the server public key, we have:
    \[
    T = F_x(H(\text{U} \parallel \text{P}))^t \cdot \text{SessionID}^x \mod q
    \]
\end{itemize}

{The use of blinding factors in both the \gls{OPRF} input and the session ID ensures that the server cannot link tokens across sessions, even if it observes multiple tokens. Each token is uniquely blinded, meaning that even if the same username and password are used, the resulting session tokens are cryptographically unlinkable.}


\item \textbf{Anonymous Session Token Verification:}  
The client sends the token \( T \), the blinded hash \( H(U \parallel P)^t \), and the session identifier \( \text{SessionID} \) to the server. Upon receiving these values, the server verifies the authenticity of the token by recomputing the expected value using its private key \( x \) and checking the following equation:
\[
T \stackrel{?}{=} (H(U \parallel P)^t)^x \cdot \text{SessionID}^x \mod q
\]
If the equation holds, it confirms that the token was correctly derived from valid credentials, completing authentication and establishing an anonymous session.

\item \textbf{Feature Engineering:}  
The client application extracts the user's live authentication features, \( \mathbf{X}_\text{Live} = (x_1, x_2, \ldots, x_n) \). Using the user's symmetric key \( S_u \), it decrypts the stored profile data, identifies the relevant features required for risk assessment, and retrieves them from the decrypted profile, forming \( \mathbf{X}_\text{Profile} = (x_1, x_2, \ldots, x_n) \).

 \item \textbf{\gls{DP} Application:}
The client application applies \gls{DP} techniques to both
 user profile features \( \mathbf{X}_\text{Profile} \) and live authentication features \( \mathbf{X}_\text{Live} \) locally before transmission to the server. This ensures that the data shared with the server does not compromise user privacy and maintains the integrity of the authentication process.
\begin{itemize}

    \item \textbf{Applying \gls{DP}:} 
    To ensure that  shared features do not reveal sensitive information, \gls{DP} is applied using the Laplace mechanism. The client application adds Laplacian noise to each feature to produce differentially private features. The Laplace mechanism is chosen because it provides strong privacy guarantees. The noise is sampled from a Laplace distribution with mean 0 and scale parameter $ b $:
 \begin{align*}
      \mathbf{X}_\text{Profile}' &= \mathbf{X}_\text{Profile} + \text{Lap}(\Delta f / \epsilon)\\
    \mathbf{X}_\text{Live}' &= \mathbf{X}_\text{Live} + \text{Lap}(\Delta f / \epsilon)   
 \end{align*}
The noise parameter \( \lambda = \frac{\Delta f}{\epsilon} \) is determined by the desired privacy budget \( \epsilon \) and the sensitivity \( \Delta f \). The sensitivity \( \Delta f \) quantifies the maximum possible change in the output of the risk evaluation function \( f \) when a single feature in the user profile or live authentication data is modified.

    \item \textbf{Transmission of Differentially Private Features:}  
    {The client application securely transmits the differentially private features} $ \mathbf{X}_\text{Profile}' $ and $ \mathbf{X}_\text{Live}' $ {to the server for risk evaluation.}
\end{itemize}

\item \textbf{Risk Score Computation and Adaptive Authentication:}
This step involves the server computing a risk score based on the differentially private features received from the user and adjusting the authentication requirements accordingly. This ensures that the authentication process is both secure and adaptive to the risk level.
\begin{itemize}
    \item \textbf{Risk Score Computation on Server:} 
    The server computes the risk score based on the received differentially private features. The server aggregates the differentially private features $ \mathbf{X}_\text{Profile}' $ and $ \mathbf{X}_\text{Live}' $.
    Using a predefined risk assessment mechanism $ f $, like the ones presented in Section \ref{risk-evaluation}, the server computes the risk score $ R $ (i.e., $ R = f(\mathbf{X}_\text{Profile}', \mathbf{X}_\text{Live}') $). The function $ f $ may involve weighted sums, \gls{ML} models, or other techniques to assess the risk level.
    
    \item \textbf{Adaptive Authentication Based on Risk Score:} 
    {
Based on the computed risk score} $ R ${, the server dynamically adjusts the authentication requirements to align with the perceived level of risk. If the risk score is low, standard authentication procedures may suffice; however, for higher risk scores, additional authentication factors may be required. This adaptive mechanism tailors the authentication process to the specific risk level, effectively balancing security with user convenience.}

\end{itemize}

\end{enumerate}





\subsection{Backup and Recovery}

\begin{enumerate}[wide, font=\bfseries, labelwidth=!, labelindent=0pt, label=\textbf{Step} \arabic*.]

\item \textbf{Backup:}  
{The client periodically updates the encrypted profile} \( E_{S_u}(\mathbf{X}_\text{Profile}) \) {in} \gls{IPFS}. {After each upload, the corresponding} \gls{CID} {is returned. To ensure recoverability, the client sends a signed transaction to a smart contract on the blockchain, binding the new CID to their persistent public identifier (e.g., public key or account address). The smart contract maintains a mapping between the user’s identity and their latest profile CID, ensuring only the rightful owner can update this mapping.}

\item \textbf{Recovery:}  
{In the event of device loss or failure, the user installs the client application on a new device and provides their public identifier. The application queries the smart contract to retrieve the latest CID associated with that identifier. Using this CID, the encrypted profile is fetched from} \gls{IPFS}{, and the user can decrypt it using their symmetric key} \( S_u \){, which must be backed up separately (e.g., via secure cloud storage or a physical backup).}
\end{enumerate}

\section{{Security and Privacy Analysis}}\label{sec:security}

This section provides a comprehensive analysis of the proposed protocol's security and privacy guarantees. We outline the underlying assumptions, define key security properties, and present formal proofs that validate the protocol's robustness against potential attacks. By examining its components under these assumptions, we demonstrate its ability to meet the necessary security and privacy requirements.

\subsection{{Security Assumptions}}

The security of the proposed protocol relies on the following assumptions:

\begin{itemize}
    \item \textbf{Security of the Encryption Scheme:} {The encryption scheme $ E $ used to encrypt user profiles is semantically secure (IND-CPA secure). This ensures that an adversary cannot distinguish between ciphertexts of any two chosen plaintexts with any non-negligible advantage.}
    
    \item \textbf{Cryptographic Hash Functions:} {The hash function $ H $ used in the protocol is modeled as a random oracle, producing independent and uniformly distributed outputs for any distinct inputs.}
    
    
    
    \item \textbf{Uniform and Independent Randomness:} {All randomness used in the protocol (e.g., for key generation, blinding factors) is assumed to be uniform and independent, including the random coin flip $ b $ in the challenge phase of the security game.}

\item \textbf{Hardness of the \gls{DDH} Problem:} Given a randomly chosen generator \( g \) of a cyclic group of prime order \( q \), and elements \( g^x \) and \( g^y \), it is computationally infeasible for an adversary to distinguish whether a third element \( g^z \) corresponds to \( g^{xy} \)  or is instead a random element in the group. 

\end{itemize}

\subsection{Security Proofs}

Building on the aforementioned assumptions, we now present formal proofs to validate the protocol's security properties.

\subsubsection{\gls{OPRF} Security}
We begin by establishing the foundational guarantees of our \gls{OPRF}: pseudorandomness and obliviousness. Pseudorandomness ensures that \gls{OPRF} outputs are indistinguishable from random, while obliviousness guarantees client privacy by preventing the server from learning user inputs.

\begin{game}[\textbf{Pseudorandomness of \gls{OPRF}}]
    We define the following game between a challenger $\mathcal{C}$ and an adversary $\mathcal{A}$:

\begin{enumerate}
    \item \textbf{Setup:} $\mathcal{C}$ selects a random $x \in \mathbb{Z}_q^*$ and defines the \gls{OPRF} function $F_x(m) = m^x \mod q$.
    
    \item \textbf{Challenge:} $\mathcal{A}$ submits polynomially many queries $m_1, \dots, m_q$ to $\mathcal{C}$. $\mathcal{C}$ selects a random bit ${\gamma_{b}}$. If ${\gamma_{b}} = 0$, $\mathcal{C}$ responds with $F_x(m_i) = m_i^x \mod q$.  If ${\gamma_{b}} = 1$, $\mathcal{C}$ responds with random values $R(m_i) \sim \mathbb{Z}_q^*$.
    \item \textbf{Adversary’s Goal:} $\mathcal{A}$ outputs a guess ${\gamma'_{b}}$ for ${\gamma_{b}}$.
\end{enumerate}
The advantage of $\mathcal{A}$ in the above game is defined as:
\[
\text{Adv}_{\text{pseudo}}^{\mathcal{A}} = \left| \Pr[{\gamma'_{b}} = {\gamma_{b}}] - \frac{1}{2} \right|.
\]
\end{game}

\begin{lem}[\textbf{Pseudorandomness of \gls{OPRF}}]\label{Pseudorandomness}
The \gls{OPRF} used in our protocol is pseudorandom under the \gls{DDH} assumption. Specifically, for any \gls{PPT} adversary $\mathcal{A}$, its advantage in distinguishing the \gls{OPRF} output from random is negligible.
\end{lem}

\begin{proof}
We prove this by reduction to the \gls{DDH} problem. Suppose, for contradiction, that there exists a \gls{PPT} adversary $\mathcal{A}$ that distinguishes the \gls{OPRF} output from a random function with non-negligible advantage. We construct a simulator $\mathcal{B}$ that, given a \gls{DDH} challenge tuple $(g, g^x, g^y, g^z)$, attempts to distinguish whether $g^z = g^{xy}$ or $g^z$ is a random group element. To do so, $\mathcal{B}$ interacts with $\mathcal{A}$ by simulating the game defined for  pseudorandomness of \gls{OPRF}. It forwards all queries $m_1, \dots, m_q$ from $\mathcal{A}$ and selects a random bit ${\gamma_{b}}$. If ${\gamma_{b}} = 0$, it responds with $F_x(m_i) = (g^{r_i})^x = (g^x)^{r_i} \mod q$ for some known $r_i \in \mathbb{Z}_q$, embedding the \gls{DDH} challenge element $g^x$. If ${\gamma_{b}} = 1$, it responds with random values $R(m_i) \sim \mathbb{Z}_q^*$.
If $\mathcal{A}$ can distinguish between these cases with non-negligible probability, then it can determine whether $g^z = g^{xy}$, thereby solving the \gls{DDH} problem. Since \gls{DDH} is assumed to be hard, this contradicts our assumption that $\mathcal{A}$ has a non-negligible advantage. Thus, $\mathcal{A}$'s advantage must be negligible, proving the pseudorandomness of the \gls{OPRF}.
\end{proof}

\begin{game}[\textbf{Obliviousness of \gls{OPRF}}]
We define the following game between a challenger $\mathcal{C}$ and an adversary $\mathcal{A}$:

\begin{enumerate}
    \item \textbf{Setup:} $\mathcal{C}$ selects a random $x \in \mathbb{Z}_q^*$.

    \item \textbf{Interaction:} $\mathcal{A}$ selects a message $m$ and a random blinding factor $b \in \mathbb{Z}_q^*$, then computes the blinded message \(
    m' = m \cdot g^b \mod q
    \) and sends it to $\mathcal{C}$.

    \item \textbf{Challenge:} $\mathcal{C}$ selects a random bit ${\gamma_{b}}$. If ${\gamma_{b}} = 0$, $\mathcal{C}$ responds with $F_x(m') = (m')^x \mod q$. If ${\gamma_{b}} = 1$, $\mathcal{C}$ responds with a random value $R(m') \sim \mathbb{Z}_q^*$.

    \item \textbf{Adversary’s Goal:} $\mathcal{A}$ outputs a guess ${\gamma'_{b}}$ for ${\gamma_{b}}$.
\end{enumerate}

The advantage of $\mathcal{A}$ in the above game is defined as:
\[
\text{Adv}_{\text{oblivious}}^{\mathcal{A}} = \left| \Pr[{\gamma'_{b}} = {\gamma_{b}}] - \frac{1}{2} \right|.
\]

\end{game}

\begin{lem}[\textbf{Obliviousness of \gls{OPRF}}]\label{lemma:oprf-oblivious}
The \gls{OPRF} used in our protocol is oblivious under the \gls{DDH} assumption. Specifically, for any \gls{PPT} adversary $\mathcal{A}$, its advantage in distinguishing the \gls{OPRF} output from random is negligible.
\end{lem}
\begin{proof}
We prove this by reduction to the \gls{DDH} problem. Suppose, for contradiction, that there exists a \gls{PPT} adversary $\mathcal{A}$ that distinguishes the \gls{OPRF} output from a random function with non-negligible advantage. We construct a simulator $\mathcal{B}$ that, given a \gls{DDH} challenge tuple $(g, g^x, g^y, g^z)$, attempts to distinguish whether $g^z = g^{xy}$ or $g^z$ is a random group element.
To do so, $\mathcal{B}$ interacts with $\mathcal{A}$ by simulating the game defined for obliviousness of \gls{OPRF}. $\mathcal{A}$ selects a message $m$ and a blinding factor $b \in \mathbb{Z}_q^*$, computes the blinded message $m' = m \cdot g^b \mod q$, and sends it to $\mathcal{B}$. 
$\mathcal{B}$ selects a random bit ${\gamma_{b}}$. If ${\gamma_{b}} = 0$, $\mathcal{B}$ sets $g^b = g^y$ (where $g^y$ is provided in the \gls{DDH} challenge) and computes
    \(    F_x(m') = (m \cdot g^b)^x = m^x \cdot g^{bx}.
    \)
    Since $g^{bx} = g^{yx}$, $\mathcal{B}$ sets the response to $g^z$ from the \gls{DDH} challenge.  If ${\gamma_{b}} = 1$, $\mathcal{B}$ responds with a uniformly random element $R(m')$.  
If $\mathcal{A}$ can distinguish between these cases with non-negligible probability, then it can determine whether $g^z = g^{xy}$, thereby solving the \gls{DDH} problem. Since \gls{DDH} is assumed to be hard, this contradicts our assumption that $\mathcal{A}$ has a non-negligible advantage. Thus, $\mathcal{A}$'s advantage must be negligible, proving the obliviousness of the \gls{OPRF}.
\end{proof}
\begin{cor}[\textbf{Credential Confidentiality via Obliviousness}]
\label{cor:credential-confidentiality}
Lemma~\ref{lemma:oprf-oblivious} directly implies that a server or any
adversary $\mathcal{A}$ observing only the blinded value $m' = m \cdot g^b$ 
\emph{cannot} recover $m$ with more than negligible probability. Under the \gls{DDH}
assumption, there is no efficient way to invert $m' = m \cdot g^b$ to learn $m$.
Thus, an honest-but-curious server fails to obtain the user’s hashed password
$m = H(P)$ beyond trivial guessing, establishing strong credential
confidentiality in our OPRF-based authentication protocol.
\begin{proof}
This follows immediately from Lemma~\ref{lemma:oprf-oblivious}, as recovering $m$ from $m' = m \cdot g^b$ would contradict the obliviousness of the \gls{OPRF}.
\end{proof}
\end{cor}

\subsubsection{Confidentiality of User Profiles}
The confidentiality of user profiles is guaranteed by the \gls{IND-CPA} security of the encryption scheme. This ensures that an adversary cannot distinguish between encryptions of different user profiles.
\begin{game}[\textbf{\gls{IND-CPA}}]
We define the standard \gls{IND-CPA} security game between a challenger \( \mathcal{C} \) and an adversary \( \mathcal{A} \):

\begin{enumerate}
    \item \textbf{Setup:}   \( \mathcal{C} \) generates a random symmetric encryption key \( S_u \).  \( \mathcal{A} \) is provided with any public parameters of the encryption scheme.

    \item \textbf{Challenge:}  \( \mathcal{A} \) selects two plaintext user profiles \( \mathbf{X}_0, \mathbf{X}_1 \) of equal length and submits them to \( \mathcal{C} \).  \( \mathcal{C} \) selects a random bit \( {\gamma_{b}} \in \{0,1\} \) and encrypts the corresponding plaintext \(
        C^* = E_{S_u}(\mathbf{X}_{{\gamma_{b}}}).
        \)  \( \mathcal{C} \) sends \( C^* \) to \( \mathcal{A} \).
    
    \item \textbf{Adversary’s Goal:}  \( \mathcal{A} \) attempts to determine which plaintext was encrypted.  \( \mathcal{A} \) outputs a guess \( {\gamma'_{b}} \).

\end{enumerate}
 The adversary wins if \( {\gamma'_{b}} = {\gamma_{b}} \) with significantly higher probability than random guessing.
The adversary's advantage in the above game  is given by:
\[
   \text{Adv}_{\text{IND-CPA}}^{\mathcal{A}} = \left| \Pr[\mathcal{A}(C^*) = {\gamma_{b}}] - \frac{1}{2} \right|.
\]
\end{game}
The confidentiality advantage of an adversary \( \mathcal{A} \) is defined as \(
\text{Adv}_{\text{conf}}^{\mathcal{A}} = \text{Adv}_{\text{IND-CPA}}^{\mathcal{A}}
\), which quantifies  \( \mathcal{A} \)'s success in distinguishing encryptions of different user profiles.

\begin{thm}[\textbf{Confidentiality of User Profiles}]
If the encryption scheme \( E \) is IND-CPA secure, then the confidentiality of encrypted user profiles is preserved. 
Formally, assuming the IND-CPA security of \( E \), for any \gls{PPT} adversary \(\mathcal{A} \) attempting to break the confidentiality of user profiles, there exists a negligible function \( \text{negl}(\lambda) \) such that
\(
\text{Adv}_{\text{conf}}^{\mathcal{A}} \leq \text{negl}(\lambda),
\)
where \( \lambda \) is the bit-length of the secret key \( S_u \) used for encryption.
\end{thm}

\begin{proof}
We prove this by a reduction to the IND-CPA security of the encryption scheme \( E \).
Suppose that there exists a \gls{PPT} adversary \( \mathcal{A} \) that can distinguish between \( E_{S_u}(\mathbf{X}_0) \) and \( E_{S_u}(\mathbf{X}_1) \) with non-negligible advantage.
We construct an adversary \( \mathcal{B} \) that uses \( \mathcal{A} \) to break the \gls{IND-CPA} security of \( E \). 
\( \mathcal{B} \) receives \( \mathbf{X}_0, \mathbf{X}_1 \)  from \( \mathcal{A} \) and forwards them to the \gls{IND-CPA} challenger.  The \gls{IND-CPA} challenger selects a random bit \( {\gamma_{b}} \) and returns the challenge ciphertext \( C^* = E_{S_u}(\mathbf{X}_{\gamma_{b}}) \).  \( \mathcal{B} \) replays \( C^* \) to \( \mathcal{A} \) and \( \mathcal{A} \) outputs a guess \( {\gamma'_{b}} \).  \( \mathcal{B} \) output \( {\gamma'_{b}} \) as its own guess for \( {\gamma_{b}} \).
Since \( \mathcal{B} \) is simulating the exact IND-CPA game for \( \mathcal{A} \), the advantage of \( \mathcal{A} \) in distinguishing the encryptions is equal to the advantage of \(\mathcal{B}\) in the IND-CPA game (i.e., \(
   \text{Adv}_{\text{conf}}^{\mathcal{A}} = \text{Adv}_{\text{IND-CPA}}^{\mathcal{B}}
\)).
Since \( E \) is IND-CPA secure, the right-hand side is negligible, contradicting our assumption. Thus, no adversary \( \mathcal{A} \) can achieve a non-negligible advantage, proving the confidentiality of encrypted user profiles.
\end{proof}

\subsubsection{Authenticity of User Tokens}
Our protocol ensures that only legitimate users can generate valid authentication tokens. This is formally proven by showing that forging a valid token without the secret key is computationally infeasible, assuming the pseudorandomness of the \gls{OPRF}. This guarantees resistance to impersonation, session hijacking, and unauthorized access.


\begin{game}[\textbf{Authenticity of User Tokens}]
The game between a challenger \( \mathcal{C} \) and an adversary \( \mathcal{A} \) proceeds as follows:

\begin{enumerate}
    \item  \textbf{Setup:} The challenger \( \mathcal{C} \) generates a secret key \( x \leftarrow \mathbb{Z}_q^* \), a public key \( y = g^x \mod q \), a session-specific random value \( t \), and defines the \gls{OPRF} function \(
    F_x(m) = m^x \mod q.
    \)

    \item \textbf{Challenge:} The adversary \( \mathcal{A} \) queries authentication tokens for arbitrary user-password pairs \( (U, P) \) and receives \(T = F_x(H(U \parallel P)) \oplus t\).
    
    \item \textbf{Adversary’s Goal:} \( \mathcal{A} \) outputs a candidate token \( T^* \) for an unqueried pair \( (U^*, P^*) \), aiming to satisfy:
    \[
    T^* = F_x(H(U^* \parallel P^*)) \oplus t^*.
    \]
    The adversary wins if it can produce a valid authentication token for an unqueried \((U^*, P^*)\).
\end{enumerate}

The adversary’s advantage in forging a valid token is defined as:
\[
\text{Adv}_{\text{auth}}^{\mathcal{A}} = \Pr[\mathcal{A} \text{ forges a valid token}].
\]
\end{game}

\begin{thm}[\textbf{Authenticity of User Tokens}]
If the \gls{OPRF} function \( F_x \) is pseudorandom (proven in Lemma~\ref{Pseudorandomness}), then for any \gls{PPT} adversary \( \mathcal{A} \), the probability of forging a valid authentication token is negligible:
\[
\text{Adv}_{\text{auth}}^{\mathcal{A}} \leq q_{\text{auth}} \cdot \text{Adv}_{\text{pseudo}}^{\mathcal{A}},
\]
where \( q_{\text{auth}} \) is the number of authentication queries made by \( \mathcal{A} \).
\end{thm}

\begin{proof}
We prove this by reduction to the pseudorandomness of the \gls{OPRF}, as established in Lemma~\ref{Pseudorandomness}. Suppose, for contradiction, that there exists a \gls{PPT} adversary \( \mathcal{A} \) that forges a valid authentication token with non-negligible probability. We construct a reduction adversary \( \mathcal{B} \) that interacts with a challenger in the pseudorandomness game of \( F_x \). The challenger either implements \( F_x(m) = m^x \mod q \) or replaces it with a uniformly random function.
The reduction adversary \( \mathcal{B} \) simulates the authentication protocol for \( \mathcal{A} \) by forwarding all authentication queries \( (U, P) \) to its pseudorandomness challenger, receiving either \( F_x(H(U \parallel P)) \) or a random value, and computing the authentication token \( T = F_x(H(U \parallel P)) \oplus t \), where \( t \) is freshly chosen for each session. Since \( \mathcal{B} \) does not know \( x \), it relies entirely on the challenger’s responses.
The adversary \( \mathcal{A} \) then attempts to forge an authentication token \( T^* \) for an unqueried pair \( (U^*, P^*) \), meaning
\(
T^* \oplus t^* = F_x(H(U^* \parallel P^*))\). Since \( (U^*, P^*) \) was not previously queried, \( \mathcal{A} \) must have computed or predicted \( F_x(H(U^* \parallel P^*)) \) without oracle access. If \( \mathcal{A} \) succeeds with non-negligible probability, then \( \mathcal{B} \) can distinguish whether its challenge oracle was computing a real \gls{OPRF} or a random function.
By the definition of pseudorandomness, the adversary’s success probability is bounded as \(
\text{Adv}_{\text{auth}}^{\mathcal{A}} \leq q_{\text{auth}} \cdot \text{Adv}_{\text{pseudo}}^{\mathcal{A}}\),
where \( q_{\text{auth}} \) is the number of authentication attempts made by \( \mathcal{A} \). Since we have previously established the pseudorandomness of \( F_x \) under the \gls{DDH} assumption, \( \mathcal{A} \) must have negligible advantage in distinguishing \( F_x(H(U^* \parallel P^*)) \) from a random value, proving the theorem.
\end{proof}

\subsubsection{Anonymity and Unlinkability} To ensure robust privacy, our protocol establishes three essential properties: token anonymity, which prevents adversaries from linking authentication tokens to specific users; token unlinkability, which ensures that tokens generated across different sessions remain uncorrelated, preventing tracking over time; and user anonymity, which guarantees that an adversary cannot infer whether a particular user is participating in the authentication process. The following proofs formally validate these properties under the defined security assumptions, demonstrating the protocol’s resilience against linking, tracking, and inference attacks.
\begin{game}[\textbf{Anonymity of User Tokens}]
We define the security game between a challenger \( \mathcal{C} \) and an adversary \( \mathcal{A} \) as follows:

\begin{enumerate}
    \item \textbf{Setup:}  
     \( \mathcal{C} \) selects two distinct users \( U_0 \) and \( U_1 \), along with their credentials \( P_0 \) and \( P_1 \).

    \item \textbf{Challenge:}  
     \( \mathcal{C} \) selects a random bit \( {\gamma_{b}} \in \{0,1\} \) and computes the authentication token \(T^* = F_x(H(U_{{\gamma_{b}}} \parallel P_{{\gamma_{b}}})) \oplus t^*\).
    \item \textbf{Adversary’s Goal:}  
    \( \mathcal{A} \) attempts to determine the value of \( {\gamma_{b}} \) by analyzing \( T^* \) and outputs a guess \( {\gamma'_{b}} \). The adversary wins if they correctly guess \( {\gamma_{b}} \) with a probability significantly greater than random guessing.

\end{enumerate}

 The advantage of \( \mathcal{A} \) in breaking anonymity is defined as:

\[
\text{Adv}_{\text{anon}_t}^{\mathcal{A}} = \left| \Pr[{\gamma'_{b}} = {\gamma_{b}}] - \frac{1}{2} \right|.
\]

\end{game}

\begin{thm}[\textbf{Anonymity of User Tokens}]
If the \gls{OPRF} function \( F_x \) is pseudorandom (proven in Lemma~\ref{Pseudorandomness}), then for any \gls{PPT} adversary \( \mathcal{A} \), the probability of linking a token to a user is negligible:
\[
\text{Adv}_{\text{anon}_t}^{\mathcal{A}} \leq q_h \cdot \text{Adv}_{\text{pseudo}}^{\mathcal{B}}.
\]
where \( q_h \) is the number of authentication token queries made by \( \mathcal{A} \).
\end{thm}
\begin{proof}
Suppose, for contradiction, that there exists a PPT adversary \( \mathcal{A} \) that can link a user token \( T^* \) to its corresponding user with non-negligible advantage \( \text{Adv}_{\text{anon}_t}^{\mathcal{A}} \). We construct a reduction adversary \( \mathcal{B} \) that interacts with an OPRF pseudorandomness challenger \( \mathcal{C} \), which implements either the real OPRF \( F_x(m) = m^x \bmod q \) or a random function \( R(m) \sim \mathbb{Z}_q^* \). The reduction adversary \( \mathcal{B} \) simulates the authentication protocol for \( \mathcal{A} \), responding to authentication queries using its oracle \( \mathcal{O}(\cdot) \), where it computes authentication tokens as \( T = \mathcal{O}(H(U \parallel P)) \oplus t \). During the challenge phase, \( \mathcal{A} \) selects two users \( (U_0, P_0) \) and \( (U_1, P_1) \), and \( \mathcal{B} \) provides \( T^* = \mathcal{O}(H(U_{\gamma_b} \parallel P_{\gamma_b})) \oplus t^* \), where \( \gamma_b \) is a hidden bit. If \( \mathcal{A} \) correctly distinguishes \( T^* \), then \( \mathcal{B} \) outputs the same guess in the OPRF pseudorandomness game, meaning \( \mathcal{B} \) successfully breaks the pseudorandomness of the OPRF. Since \( \mathcal{A} \) makes at most \( q_h \) authentication queries, a union bound gives \( \text{Adv}_{\text{anon}_t}^{\mathcal{A}} \leq q_h \cdot \text{Adv}_{\text{pseudo}}^{\mathcal{B}} \). Under the \gls{DDH} assumption, the OPRF is pseudorandom, meaning \( \text{Adv}_{\text{pseudo}}^{\mathcal{B}} = negl(\lambda) \), and thus, \( \text{Adv}_{\text{anon}_t}^{\mathcal{A}} \) is also negligible, proving the theorem.
\end{proof}

\begin{game}[\textbf{Unlinkability of User Tokens}]
We define the security game between a challenger \( \mathcal{C} \) and an adversary \( \mathcal{A} \):

\begin{enumerate}
    \item \textbf{Setup:} \( \mathcal{C} \) selects a random OPRF key \( x \in \mathbb{Z}_q^* \) and defines the function \( F_x(m) = m^x \mod q \). For each authentication session, \( \mathcal{C} \) generates tokens using fresh randomness.

    \item \textbf{Challenge:} \( \mathcal{C} \) selects a random bit \( \gamma_b \in \{0,1\} \). If \( \gamma_b = 0 \), both tokens \( (T_0, T_1) \) are generated from the same user in different sessions. If \( \gamma_b = 1 \), the tokens correspond to different users. The adversary \( \mathcal{A} \) receives \( (T_0, T_1) \).

    \item \textbf{Adversary’s Goal:} \( \mathcal{A} \) attempts to guess \( \gamma_b \), outputting \( \gamma'_b \). The adversary wins if \( \gamma'_b = \gamma_b \) with a probability significantly greater than \( \frac{1}{2} \).
\end{enumerate}

The adversary’s advantage in breaking unlinkability is defined as:
\[
\text{Adv}_{\text{unlink}}^{\mathcal{A}} = \left| \Pr[{\gamma'_{b}} = {\gamma_{b}}] - \frac{1}{2} \right|.
\]
\end{game}

\begin{thm}[\textbf{Unlinkability of User Tokens}]
Assuming the \gls{OPRF} function \( F_x \) is pseudorandom (as proven in Lemma~\ref{Pseudorandomness}), the adversary’s advantage in linking authentication tokens across sessions is negligible:
\[
\text{Adv}_{\text{unlink}}^{\mathcal{A}} \leq q_h \cdot \text{Adv}_{\text{pseudo}}^{\mathcal{B}},
\]
where \( q_h \) is the number of authentication queries and \( \text{Adv}_{\text{pseudo}}^{\mathcal{B}} \) is the advantage of a PPT adversary \( \mathcal{B} \) in distinguishing the OPRF from a random function.
\end{thm}

\begin{proof}
Suppose that there exists an adversary \( \mathcal{A} \) that wins the unlinkability game with a non-negligible advantage \( \text{Adv}_{\text{unlink}}^{\mathcal{A}} \). We construct a reduction adversary \( \mathcal{B} \) that interacts with an OPRF pseudorandomness challenger \( \mathcal{C} \), which implements either the real OPRF \( F_x(m) = m^x \bmod q \) or a random function \( R(m) \sim \mathbb{Z}_q^* \).  \( \mathcal{B} \) simulates the authentication system for \( \mathcal{A} \), using its oracle \( \mathcal{O}(\cdot) \) to calculate the authentication tokens as \( T = \mathcal{O}(H(U \parallel P)) \oplus t \), where \( t \) is session randomness. During the unlinkability challenge, \( \mathcal{B} \) selects two user credentials \( (U_0, P_0) \) and \( (U_1, P_1) \) and provides the challenge tokens \( (T_0, T_1) \), where \( T_0 = \mathcal{O}(H(U_{\gamma_b} \parallel P_{\gamma_b})) \oplus t_0 \) and \( T_1 = \mathcal{O}(H(U_{\gamma_b} \parallel P_{\gamma_b})) \oplus t_1 \) if \( \gamma_b = 0 \), or from different users if \( \gamma_b = 1 \). 
If \( \mathcal{A} \) distinguishes whether the tokens belong to the same or different users with non-negligible advantage \( \varepsilon \), then \( \mathcal{B} \) distinguishes whether its oracle implements a real OPRF or a random function with the same advantage \( \varepsilon \), contradicting the pseudorandomness of the OPRF. Since \( \mathcal{A} \) makes at most \( q_h \) queries, a union bound gives \( \text{Adv}_{\text{unlink}}^{\mathcal{A}} \leq q_h \cdot \text{Adv}_{\text{pseudo}}^{\mathcal{B}} \). Under the \gls{DDH} assumption, the OPRF is pseudorandom, meaning \( \text{Adv}_{\text{pseudo}}^{\mathcal{B}} = {negl}(\lambda) \), which implies \( \text{Adv}_{\text{unlink}}^{\mathcal{A}} \) is negligible, proving the theorem.
\end{proof}
\begin{game}[\textbf{Anonymity of User}]
We define a security game between a challenger \( \mathcal{C} \) and an adversary \( \mathcal{A} \):

\begin{enumerate}
    \item \textbf{Setup:} \( \mathcal{C} \) generates the server’s private key \( x \) and sets up the \gls{OPRF} function \( F_x(m) = m^x \mod q \).  \( \mathcal{A} \) is given access to a set of valid authentication tokens \( \{T_i\} \) generated from multiple users.

    \item \textbf{Challenge:}  \( \mathcal{A} \) selects two users \( U_0 \) and \( U_1 \) and submits them to \( \mathcal{C} \).  \( \mathcal{C} \) selects a random bit \( {\gamma_{b}} \in \{0,1\} \) and generates a challenge token \( T^* \) using \( U_{\gamma_{b}} \)'s credentials.  \( \mathcal{C} \) sends \( T^* \) to \( \mathcal{A} \).

    \item \textbf{Adversary’s Goal:} \( \mathcal{A} \) attempts to determine whether \( T^* \) belongs to \( U_0 \) or \( U_1 \).  \( \mathcal{A} \) outputs a guess \( {\gamma'_{b}} \) for \( {\gamma_{b}} \).  The adversary wins if \( {\gamma'_{b}} = {\gamma_{b}} \) with significantly higher probability than random guessing.
\end{enumerate}

The adversary’s advantage in breaking user anonymity is defined as:
\[
   \text{Adv}_{\text{anon}_u}(\mathcal{A}) = \left| \Pr[\mathcal{A}(T^*) = {\gamma_{b}}] - \frac{1}{2} \right|.
\]
\end{game}

\begin{thm}[\textbf{Anonymity of User}]
Assuming the \gls{OPRF} function is pseudorandom (IND-CPA secure), 
the proposed protocol ensures user anonymity. Specifically, the advantage 
of any \gls{PPT} adversary \( \mathcal{A} \) in linking authentication 
tokens to users is negligible.
\end{thm}

\begin{proof}
Suppose there exists a \gls{PPT} adversary \( \mathcal{A} \) that can 
distinguish whether a given authentication token \( T^* \) belongs to 
\( U_0 \) or \( U_1 \) with non-negligible advantage.
We construct a simulator \( \mathcal{B} \) that uses \( \mathcal{A} \) 
to break the pseudorandomness of the \gls{OPRF}. To do that, first  \( \mathcal{B} \) receives a challenge \gls{OPRF} output \( F_x(m) \) from an OPRF pseudorandomness challenger. \( \mathcal{B} \) simulates the anonymity game, generating 
    authentication tokens and executing the protocol as \( \mathcal{C} \).  \( \mathcal{B} \) sends the challenge token \( T^* \) 
    (computed using \( F_x(m) \)) to \( \mathcal{A} \). Then  \( \mathcal{A} \) outputs a guess \( {\gamma'_{b}} \). At the end  \( \mathcal{B} \) outputs \( {\gamma'_{b}} \) as its own guess for whether the given \gls{OPRF} output is real or random.
If \( \mathcal{A} \) wins the anonymity game with non-negligible probability, 
then \( \mathcal{B} \) successfully distinguishes real \gls{OPRF} outputs from random ones, breaking the pseudorandomness of the \gls{OPRF}. This contradicts the assumed OPRF pseudorandomness challenger. Thus, the 
adversary's advantage \( \text{Adv}_{\text{anon}_u}(\mathcal{A}) \) is 
negligible, proving that user anonymity is preserved.
\end{proof}

\subsubsection{Differential Privacy in Risk Assessment}
To enhance privacy and prevent profiling, inference, and correlation attacks while enabling adaptive authentication, our protocol uses \gls{DP}. \gls{DP} ensures that authentication decisions do not reveal sensitive user attributes by introducing calibrated noise, limiting the impact of individual features on the calculated risk score. By setting the noise parameter \(\lambda\) appropriately, we maintain similar probability distributions for adjacent datasets, mitigating tracking risks while preserving meaningful authentication. The following security game and proof formally establish these guarantees.
\begin{game}[\textbf{Differential Privacy}]\label{DP-game-specific}
We define a privacy game between a challenger \( \mathcal{C} \) and an adversary \( \mathcal{A} \):

\begin{enumerate}
    \item \textbf{Setup:} \( \mathcal{C} \) samples two adjacent datasets \( D = (\mathbf{X}_\text{Profile}, \mathbf{X}_\text{Live}) \) and  
          \( D' = (\mathbf{X}_\text{Profile}', \mathbf{X}_\text{Live}') \), differing by at most one feature.
    \item  \textbf{Challenge:} \( \mathcal{C} \) selects a random bit \( \gamma_{b} \in \{0,1\} \). If \( \gamma_{b} = 0 \), the risk-based authentication mechanism is executed on \( D \); otherwise, it is executed on \( D' \). 
    \item \textbf{Guess:} The adversary \( \mathcal{A} \) receives the risk score \( R \) and attempts to distinguish whether it was computed on \( D \) or \( D' \).  \( \mathcal{A} \) outputs a guess \( \gamma'_{b} \) for \( \gamma_{b} \).
\end{enumerate}

The adversary's advantage in this game is defined as:
\[
\text{Adv}_{\mathcal{A}} = \left| \Pr[\gamma'_{b} = \gamma_{b}] - \frac{1}{2} \right|.
\]

\end{game}

\begin{thm}[\textbf{Differential Privacy}]
The risk scoring mechanism \( f \) in the proposed protocol satisfies \( \epsilon \)-differential privacy, ensuring that an individual feature has a limited effect on the computed risk score.
\end{thm}

\begin{proof}
Suppose there exists an adversary \( \mathcal{A} \) that can distinguish between \( f(D) \) and \( f(D') \) with non-negligible advantage. We show how to use \( \mathcal{A} \) to construct another adversary \( \mathcal{B} \) that can break the differential privacy guarantee of the Laplace Mechanism. 
Since the risk function \( f \) is computed on the differentially private features \( \mathbf{X}_\text{Profile}' \) and \( \mathbf{X}_\text{Live}' \), its sensitivity is at most \( \Delta f \). Each feature is perturbed using the Laplace distribution:
\[
\text{Lap}(\lambda) = \frac{1}{2\lambda} \exp\left(-\frac{|x|}{\lambda}\right), \quad \text{where } \lambda = \frac{\Delta f}{\epsilon}.
\]
By definition of the Laplace Mechanism, for any two adjacent datasets \( D, D' \), the ratio of their output distributions is:
\begin{align*}
        \frac{\Pr[\mathcal{M}(D) = x]}{\Pr[\mathcal{M}(D') = x]}
        &= \frac{\frac{1}{2\lambda} \exp\left(-\frac{|x - f(D)|}{\lambda} \right)}
               {\frac{1}{2\lambda} \exp\left(-\frac{|x - f(D')|}{\lambda} \right)}\\
               &= \exp\left( \frac{|x - f(D')| - |x - f(D)|}{\lambda} \right)\\
               &\leq \exp\left( \frac{|f(D') - f(D)|}{\lambda} \right)\\
               &\leq \exp\left(\frac{\Delta f}{\lambda} \right).
\end{align*}
Since \( \lambda = \frac{\Delta f}{\epsilon} \), we obtain \(
\exp\left(\frac{\Delta f}{\lambda} \right) = \exp(\epsilon).
\) Thus, for any measurable set \( S \), we conclude:
\[
\Pr[\mathcal{M}(D) \in S] \leq e^\epsilon \Pr[\mathcal{M}(D') \in S].
\]
Suppose, for contradiction, that there exists an adversary \( \mathcal{A} \) with non-negligible advantage in distinguishing \( D \) from \( D' \) based on the risk score \( R \). We construct an adversary \( \mathcal{B} \) that breaks the differential privacy guarantee of the Laplace Mechanism. 
To achieve this, first \( \mathcal{B} \) receives a noisy output \( R \) of the risk scoring mechanism \( \mathcal{M} \) and must distinguish whether it was computed on \( D \) or \( D' \).  \( \mathcal{B} \)  runs \( \mathcal{A} \) on \( R \) and output \( \mathcal{A} \)’s guess \( \gamma'_{b} \).  If \( \mathcal{A} \) has a non-negligible advantage in distinguishing \( D \) from \( D' \), then \( \mathcal{B} \) can distinguish outputs from the Laplace Mechanism with non-negligible advantage, contradicting the known privacy guarantee of the Laplace Mechanism.
Thus, \( \mathcal{A} \) must have at most negligible advantage in distinguishing \( D \) from \( D' \), which completes the proof.
\end{proof}

\subsection{Summary} Our  analysis demonstrates that the proposed protocol ensures pseudorandomness and obliviousness of the \gls{OPRF} under \gls{DDH}, provides confidentiality of user profiles via 
\gls{IND-CPA}-secure encryption, guarantees authenticity of user tokens by bounding the forging probability, and achieves token anonymity, token unlinkability, and user anonymity through pseudorandom function arguments. Additionally, differential privacy protects risk-based decisions by preventing profiling and re-identification. Collectively, these properties ensure the protocol’s resilience to a wide range of attacks while preserving user privacy.

\section{Performance Evaluation}\label{sec:performance}

{In this section, we  evaluate the performance of the proposed protocol by analyzing the computational and communication overheads associated with both the setup and authentication phases. Our performance analysis demonstrates that the protocol incurs manageable overheads, making it well-suited for practical deployment. The following subsections provide a detailed examination of the protocol's efficiency in terms of computation, communication, and resilience under varying conditions.}

\subsection{Complexity Analysis}
This section presents a comprehensive analysis of the computational and communication complexity of our proposed protocol. We provide detailed assessments of both the setup and authentication phases, examining the cryptographic operations performed and the data exchanged between participating entities.

The computational complexity is quantified by the number and type of cryptographic operations required in each phase.
The primary cryptographic operations considered are symmetric encryption/decryption, elliptic curve operations, hash functions, and \gls{PRF} evaluations. During the setup phase, on the client side, the main operations include generating a symmetric key $ S_u $ and encrypting the profile features using \gls{AES}-256 encryption: $ E_{S_u}(\mathbf{X}_\text{Profile}) = \text{\gls{AES}}(S_u, \mathbf{X}_\text{Profile}) $. On the server side, the primary computational overhead involves generating an asymmetric key pair using \gls{ECDH}, where the private key is $ x $ and the public key is $ y = g^x $. In the authentication phase, client-side operations include generating a hash value $ H(U \parallel P) $, computing the blinded input for the \gls{OPRF} $ m' = H(U \parallel P) \cdot g^b $, decrypting the profile data using \gls{AES}-256 decryption $ D_{S_u}(E_{S_u}(\mathbf{X}_\text{Profile})) $, and adding Laplace noise to each feature for \gls{DP}: $ \mathbf{X}_\text{Profile}' = \mathbf{X}_\text{Profile} + \text{Lap}\left(\frac{\Delta f}{\epsilon}\right) $ and $\mathbf{X}_\text{Live}' = \mathbf{X}_\text{Live} + \text{Lap}(\Delta f / \epsilon)$. Server-side operations include evaluating the \gls{OPRF} $ F_x(m') = (m')^x $, generating and verifying the session token $ T $, aggregating differentially private features, and computing the risk score $ R = f(\mathbf{X}_\text{Profile}', \mathbf{X}_\text{Live}') $. A summary of these computational overheads is presented in Table \ref{tab:computational_overhead}.

\begin{table*}[!ht]
\centering
\footnotesize
\caption{Computational Overhead}
\resizebox{\linewidth}{!}{\begin{tabular}{|l|l|p{0.22\linewidth}|p{0.22\linewidth}|p{0.24\linewidth}|}
\hline
\textbf{Phase} & \textbf{Subphase} & \textbf{Client Application} & \textbf{Server} & \textbf{Description} \\
\hline
Setup & Key Generation & 1 Symmetric Key Generation & 1 \gls{ECDH} Key Generation & Generation of keys for encryption \\ \cline{2-5}
  & Feature Extraction & $n_f$ Feature Extractions & None & Extracting features from user profile \\ \cline{3-5}
 &   & 1 Symmetric Encryption & None & Encrypting the entire user profile \\  
\hline
Authentication & Token Generation (OPRF) & 1 Hash Function, 1 \gls{PRF} Evaluation & 1 \gls{PRF} Evaluation & Generating and evaluating \gls{OPRF} \\ \cline{2-5}
  & Token Verification & None & 1 \gls{ECDH} Signature Verification & Verifying the authenticity of tokens \\ \cline{2-5}
 & Feature Decryption & 1 Symmetric Decryption & None & Decrypting the entire user profile \\ \cline{2-5}
 & \gls{DP} & $2 \times n_f$ Noise Additions & None & Adding Laplace noise for \gls{DP} \\ \cline{2-5}
 & Risk Score Computation & None & $n_f$ Feature Aggregations, 1 Risk Score Computation & Calculating risk score from features \\ \cline{2-5}
 & Adaptive Authentication & None & 1 Decision Operation & Making authentication decisions based on risk score \\ \cline{2-5}
\hline
\end{tabular}}
\label{tab:computational_overhead}
\end{table*}
\begin{table*}[!ht]
\centering
\footnotesize
\caption{Communication Overhead}
\begin{tabular}{|l|l|l|p{0.41\textwidth}|}
\hline
\textbf{Phase} & \textbf{Data Exchanged (Client Application $\leftrightarrow$ Server)} & \textbf{Size (bytes)} & \textbf{Description} \\
\hline
Setup & Public Key & 32 bytes & Server's public key size (256 bits for \gls{ECDH}) \\
\hline
Authentication & Hash Value & 32 bytes & Size of hash value (256 bits) \\ \cline{2-4}
 & Token & 32 bytes & Size of anonymous token (256 bits) \\ \cline{2-4}
 & Encrypted Token & 32 bytes & Size of encrypted token (256 bits) \\ \cline{2-4}
 & Differentially Private Features & $2 \times 32 \times n_f$  & $n_f$ profile features and $n_f$ live features of size 256 bits each \\ \cline{2-4}
 & Risk Score & 4 bytes & Size of the risk score (32 bits) \\ \cline{2-4}
 & Authentication Adjustment & 4 bytes & Size of the adjustment decision (32 bits) \\ 
\hline
\end{tabular}
\label{tab:communication_overhead}
\end{table*}

The communication complexity measures the volume of data exchanged between the client application and the server during the setup and authentication phases. The data size is measured in bytes and includes the size of encrypted features, tokens, and differentially private features. During the setup phase, the communication overhead involves the transmission of the server's public key, which is 32 bytes for an \gls{ECDH} public key (256 bits). In the authentication phase, the communication overhead includes the exchange of the blinded input $ m' $ (256 bits), the session token $ T $ (256 bits), and the differentially private features $ \mathbf{X}_\text{Profile}' $ and $ \mathbf{X}_\text{Live}' $. For $ n_f $ features of 256 bits each, the total size is $ 2 \times 32 \times n_f $ bytes. {Additionally, the transmission of the risk score and authentication adjustment decision adds 64 bits to the total communication overhead, with each contributing 32 bits. A detailed breakdown of these communication overheads is provided in Table} \ref{tab:communication_overhead}.
\subsection{Experimental Results}
We conducted a performance evaluation using   an Apple MacBook with an M2 processor and 8 GB RAM running macOS 14.5, where both the client application and server were executed in a controlled environment. Cryptographic operations, including symmetric (\gls{AES}-256) and asymmetric (\gls{ECDH}) encryption, were benchmarked using the \texttt{cryptography} library in Python, with custom implementations for \gls{OPRF} and \gls{DP} mechanisms to accurately simulate the protocol's unique requirements.

\subsubsection{Simulated Network Conditions}  
To emulate real-world conditions, we developed a custom Python-based simulation introducing variable network delays (10–100 ms, uniform distribution) and bandwidth fluctuations via a congestion factor ranging from 0.5 to 2.0. Lower factors represented higher congestion, reducing transmission chunk sizes, increasing delays, and elevating packet loss, while higher values simulated smoother transmission with larger chunks. A 10\% packet loss probability was modeled, dynamically influenced by congestion, triggering TCP retransmissions based on \gls{RTT}. TCP was used as the transport protocol to reflect its reliability and built-in loss recovery mechanisms. Each simulation was executed 10 times, with results averaged to assess typical protocol performance.

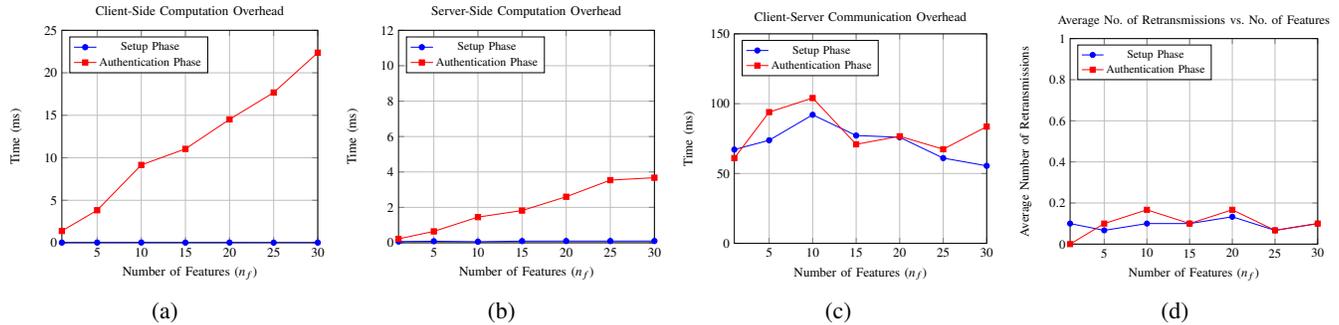
\begin{figure*}[ht]
\centering
\begin{subfigure}{0.24\textwidth}
\centering
\resizebox{\linewidth}{!}{
    \begin{tikzpicture}
            \begin{axis}[
                title={Client-Side Computation Overhead},
                xlabel={Number of Features ($n_f$)},
                ylabel={Time (ms)},
                xmin=1, xmax=30,
                ymin=0, ymax=25,
                legend pos=north west,
                grid=major,
                legend style={font=\small},
                yticklabel style={/pgf/number format/fixed,/pgf/number format/precision=3},
            ]
        \addplot[color=blue, mark=*] coordinates {
            (1, 0.020)   (5, 0.025)  (10, 0.030) (15, 0.018)(20, 0.026)     (25, 0.014)(30, 0.020)
        };
        \addlegendentry{Setup Phase}
        \addplot[color=red, mark=square*] coordinates {
            (1, 1.387)   (5, 3.828)  (10, 9.159) (15, 11.044) (20, 14.518)  (25, 17.680) (30, 22.366)
        };
        \addlegendentry{Authentication Phase}
        \end{axis}
    \end{tikzpicture}}
    \caption{}
\end{subfigure}
\begin{subfigure}{0.24\textwidth}
\centering
\resizebox{\linewidth}{!}{
    \begin{tikzpicture}
        \begin{axis}[
            title={Server-Side Computation Overhead},
            xlabel={Number of Features ($n_f$)},
            ylabel={Time (ms)},
            xmin=1, xmax=30,
            ymin=0, ymax=12,
            legend pos=north west,
            grid=major,
            legend style={font=\small},
            yticklabel style={/pgf/number format/fixed,/pgf/number format/precision=3},
        ]
        \addplot[color=blue, mark=*] coordinates {
             (1, 0.066*1.1^rand)  (5, 0.080*1.1^rand)  (10, 0.061*1.1^rand)(15, 0.090*1.1^rand) (20, 0.098*1.1^rand) 
            (25, 0.099*1.1^rand)  (30, 0.101*1.1^rand)
        };
        \addlegendentry{Setup Phase}
        \addplot[color=red, mark=square*] coordinates {
            (1, 0.215*1.1^rand)  (5, 0.666*1.1^rand)  (10, 1.328*1.1^rand) (15, 1.950*1.1^rand) (20, 2.631*1.1^rand) 
            (25, 3.298*1.1^rand) (30, 3.962*1.1^rand)
        };
        \addlegendentry{Authentication Phase}
        \end{axis}
    \end{tikzpicture}}
    \caption{}\end{subfigure}
\begin{subfigure}{0.24\textwidth}
\centering
\resizebox{\linewidth}{!}{
\begin{tikzpicture}
\begin{axis}[
    title={Client-Server Communication Overhead},
    xlabel={Number of Features ($n_f$)},
    ylabel={Time (ms)},
    xmin=1, xmax=30,
    ymin=0, ymax=150,
    legend pos=north west,
    grid=major,                 legend style={font=\small},
]
\addplot[color=blue, mark=*] coordinates {
(1, 67.136)
(5, 73.801)
(10, 92.048)
(15, 77.200)
(20, 75.942)
(25, 61.028)
(30, 55.505)
};
\addlegendentry{Setup Phase}
\addplot[color=red, mark=square*] coordinates {
(1, 61.030)
(5, 93.949)
(10, 104.040)
(15, 70.897)
(20, 76.656)
(25, 67.376)
(30, 83.571)
};
\addlegendentry{Authentication Phase}
\end{axis}
\end{tikzpicture}
}
\caption{}
\end{subfigure}
\begin{subfigure}{0.24\textwidth}
\centering
\resizebox{\linewidth}{!}{
\begin{tikzpicture}
\begin{axis}[
    title={Average No. of Retransmissions vs. No. of Features},
    xlabel={Number of Features ($n_f$)},
    ylabel={Average Number of Retransmissions},
    xmin=1, xmax=30,
    ymin=0, ymax=1,
    legend pos=north west,
    grid=major,                 legend style={font=\small},
]
\addplot[color=blue, mark=*] coordinates {
(1, 0.100)
(5, 0.067)
(10, 0.100)
(15, 0.100)
(20, 0.133)
(25, 0.067)
(30, 0.100)

};
\addlegendentry{Setup Phase}
\addplot[color=red, mark=square*] coordinates {
(1, 0.000)
(5, 0.100)
(10, 0.167)
(15, 0.100)
(20, 0.167)
(25, 0.067)
(30, 0.100)

};
\addlegendentry{Authentication Phase}
\end{axis}
\end{tikzpicture}
}
\caption{}
\end{subfigure}
\caption{{Performance Evaluation: (a) Client-Side Computation Overhead, (b) Server-Side Computation Overhead, (c) Total Communication Time, (d)  Average Number of Retransmissions}}
   \vspace{-0.4cm}
\label{fig:performance_evaluation}
\end{figure*}

\subsubsection{{Performance Metrics}} To evaluate the protocol's performance, we measured three key metrics: (a) computational overhead, which quantifies the time required for secure connection establishment (setup phase) and user authentication (authentication phase), including key generation, feature extraction, \gls{DP} application, and risk score computation; (b) communication overhead, which assesses the total data transmission time, factoring in delays and retransmissions; and (c) average number of retransmissions, which captures the frequency of packet retransmissions due to network-induced packet loss.


\subsubsection{Analysis and Interpretation}
{The results of our performance evaluation indicate that the proposed protocol incurs manageable computational and communication overheads, making it practical for real-world implementation.

\textbf{Computation Overhead:} On the client side, the setup phase overhead is minimal, primarily involving symmetric key generation and encryption. The authentication phase, while slightly more intensive due to the \gls{OPRF} and \gls{DP} operations, remains efficient, as shown in Figure} \ref{fig:performance_evaluation}.(a). {On the server side, the setup phase is dominated by \gls{ECDH} key generation, which, although computationally intensive, occurs infrequently. The authentication phase is also efficient, with primary operations such as \gls{OPRF} evaluations, token generation/verification, and risk score computation being well-optimized, as depicted in Figure} \ref{fig:performance_evaluation}.(b).

{\textbf{Communication Overhead:} As illustrated in Figure} \ref{fig:performance_evaluation}.(c){, the communication time during both the setup and authentication phases shows some variability as the number of features increases. The setup phase, involving the transmission of critical data such as public keys and initial tokens, exhibits relatively stable communication time across different feature counts. This stability is due to the fixed size of these transmissions, which do not significantly vary with the number of features.
In contrast, the authentication phase exhibits some fluctuations in communication time as the number of features increases. This is primarily due to the transmission of differentially private features and authentication tokens, which grow with the number of features. However, the overall increase in communication time is not strictly linear, likely due to the small size of each feature and the protocol's ability to keep data packets within manageable sizes. This helps minimize errors and retransmissions, demonstrating the protocol's efficiency in handling feature scaling without substantial increases in communication time.

\textbf{Average Number of Retransmissions:} The number of TCP retransmissions is closely tied to network conditions and the data transmission process. As illustrated in Figure} \ref{fig:performance_evaluation}{.(d), the number of TCP retransmissions remains relatively stable, with only minor fluctuations as the number of features increases. This stability is primarily due to the consistent size of transmissions during the setup phase and the relatively small size of individual features during the authentication phase. These factors help maintain data packets within a manageable size, reducing the likelihood of errors and subsequent retransmissions. Consequently, the proposed protocol demonstrates resilience to variations in the number of features, maintaining consistent communication performance without a substantial increase in retransmissions, even under typical network conditions.

Overall, our evaluation confirms that the proposed protocol maintains a low computational footprint and stable communication overhead, making it well-suited for real-world deployment. Its use of lightweight cryptographic primitives and differential privacy ensures robust security while preserving usability and scalability. Additionally, the protocol demonstrates resilience to network variations, ensuring adaptability across diverse environments and maintaining efficiency even as feature sets grow.



\section{Conclusion}\label{sec:conclusion}
This paper presented a privacy-enhancing protocol for risk-based adaptive authentication, addressing key challenges in secure user profiling and re-identification prevention. The proposed mechanism augments existing RBA frameworks with strong privacy guarantees, including  data confidentiality, unlinkability, and resilience against profiling attacks.
The protocol supports dynamic, risk-driven authentication with minimal usability impact. 
Formal security analysis, including proofs for OPRF pseudorandomness and obliviousness, user profile confidentiality, user token authenticity, and user anonymity and unlinkability, confirms its robustness under standard cryptographic assumptions. 
Performance evaluations demonstrate low overhead, efficient scalability to high-dimensional feature sets, and compatibility with existing identity infrastructures.
Decentralized storage via IPFS mitigates central aggregation risks and facilitates compliance with regulations such as GDPR and CCPA. Future work includes optimizing privacy-utility trade-offs under tighter DP constraints and incorporating advanced primitives like homomorphic encryption or secure multi-party computation to improve scalability and robustness.
In summary, the proposed protocol offers a scalable, efficient, and regulation-compliant enhancement to risk-based authentication, bridging adaptive security and user-centric privacy in modern digital ecosystems.



\bibliographystyle{IEEEtran}
\bibliography{bibliography}

\end{document}